\documentclass[12pt]{article}

\usepackage{amsmath}
\usepackage{amssymb}
\usepackage{amsfonts}
\usepackage{graphicx}
\usepackage{soul}

\newcommand{\rf}[1]{(\ref{#1})}
\newcommand{\beq}{\begin{equation}}
\newcommand{\beql}[1]{\beq\label{#1}}
\newcommand{\eeq}{\end{equation}}
\newcommand{\bea}{\begin{eqnarray}}
\newcommand{\eea}{\end{eqnarray}}

\newcommand{\e}{\mbox{e}}

\usepackage{color}

\newcommand{\Tr}{\mathrm{Tr}}
\newcommand{\const}{\mathrm{const}}

\newcommand{\cC}{\mathcal{C}}

\newcommand{\cN}{\mathcal{N}}

\newcommand{\cT}{\mathcal{T}}

\begin{document}

${ }$
\begin{center}
\vspace{44pt}
{ \Large \bf Characteristics of the new phase in CDT}

\vspace{30pt}

{\sl J. Ambj\o rn}$\,^{a,c}$, 
{\sl J. Gizbert-Studnicki}$\,^b$, {\sl A. G\"{o}rlich}$\,^{a,b}$,
{\sl J. Jurkiewicz}$\,^{b}$, \\
{\sl N. Klitgaard}$\,^c$ and {\sl R. Loll}$\,^{c}$

\vspace{24pt}

$^a$~The Niels Bohr Institute, Copenhagen University\\
Blegdamsvej 17, DK-2100 Copenhagen \O , Denmark.

\vspace{8pt}

$^b$~Institute of Physics, Jagiellonian University,\\
ul.~prof. Stanislawa Lojasiewicza 11,  PL 30-348 Krakow, Poland. 

\vspace{8pt}

$^c$~Institute for Mathematics, Astrophysics and Particle Physics (IMAPP)\\ 
Radboud University,
Heyendaalseweg 135, 6525 AJ Nijmegen, \\The Netherlands.

\end{center}

\vspace{30pt}

\begin{center}
{\bf Abstract}
\end{center}

\vspace{6pt}

\noindent 
Causal Dynamical Triangulations (CDT), a candidate theory of nonperturbative quantum gravity in 4D,
turns out to have a rich phase structure.
We investigate the recently discovered bifurcation phase $C_{b}$ and relate some of
its characteristics to the presence of singular vertices of very high order. The transition lines separating this phase from 
the ``time-collapsed" $B$-phase and the de Sitter phase $C_{dS}$ are of great interest when searching for physical scaling limits. 
The work presented here sheds light on the mechanisms behind these transitions. 

First, we study how the $B$-$C_{b}$ transition signal depends on the volume-fixing implemented in the simulations,
and find results compatible with the previously determined second-order character of the transition. 
The transition persists in a transfer matrix formulation, where the system's time extension is taken to be minimal. 
Second, we relate the new $C_{b}$-$C_{dS}$ transition to the appearance of singular vertices, which leads to a direct physical
interpretation in terms of a breaking of the homogeneity and isotropy observed in the de Sitter phase when crossing from $C_{dS}$
to the bifurcation phase $C_{b}$.

\vfill

\noindent
--------------------------------------------------------------------------------

{\footnotesize
\noindent
{\it emails:} ambjorn@nbi.dk, jakub.gizbert-studnicki@uj.edu.pl,  
andrzej.goerlich@uj.edu.pl, \\
jerzy.jurkiewicz@uj.edu.pl, n.klitgaard@science.ru.nl, r.loll@science.ru.nl
}

\newpage

\section{Introduction}\label{intro}

The asymptotic safety program is an attempt to describe quantum gravity as an ordinary quantum field theory. 
To overcome the well-known nonrenormalizability of the perturbative quantization, the program needs to assume the 
existence of a nonperturbative fixed point in the ultraviolet (UV). Concrete continuum calculations 
using the so-called functional renormalization group equations lend support to this assumption \cite{RG},
but necessarily involve truncations. Since the reliability of these truncations is ultimately difficult to quantify,
it is important to obtain independent evidence for the existence of a UV fixed point from
alternative, nonperturbative methods.

Defining a quantum theory by using a lattice regularization is a well-tested method for obtaining nonperturbative results. 
The arguably most spectacular results of this kind have been obtained in lattice QCD,
where the underlying theory is renormalizable, but many observables cannot be
calculated by perturbative methods. 
Lattice field theories are also well suited to finding nonperturbative UV fixed points,
which typically are associated with second-order phase transitions. 
This means that the first step in a fixed point search consists in localizing phase transition points or lines in the 
space of bare coupling constants.

In nongravitational lattice field theories the lattice approximates a piece of fixed, flat background spacetime and
the lattice spacing $a$ acts as a UV cutoff. Given that in General Relativity spacetime itself becomes dynamical,
it is natural that in a corresponding lattice field theory the lattices themselves should become dynamical entities also. 
This is precisely what happens in the approach of Dynamical Triangulations (DT) 
\cite{2DT, 3DT, 4DT} and its Lorentzian counterpart, Causal Dynamical Triangulations (CDT) \cite{CDT,CDT1,CDT2}.
Curved spacetimes, which are summed over in the gravitational path integral, are represented in the lattice regularization 
by $d$-dimensional ``lattices" constructed
from elementary building blocks, $d$-dimensional simplices of lattice link length $a$, which are glued together in all 
possible ways compatible with topological and other constraints one may impose. Note that the simplices are not ``empty", but
are pieces of flat spacetime, such that by assembling them one obtains continuous, piecewise flat manifolds,
the said triangulations.
The working hypothesis is that in the limit as $a \to 0$ this set of piecewise linear geometries becomes dense in the set of all 
continuous geometries, assuming a suitable metric can be defined on the latter.

We focus on the CDT rather than the DT approach to nonperturbative quantum gravity, because only in the CDT case 
one has observed a second-order phase transition which potentially can be used to obtain a UV scaling limit of the 
lattice theory.\footnote{A phase transition observed in DT was originally thought to be second order \cite{4DT}, 
but subsequently shown to be first order \cite{firstorder}. Recent attempts to enlarge the coupling constant space of 
DT in search of second-order transition points have so far not been successful \cite{EDT}.}
More than that, considering its conceptual simplicity and simple action (see eq.\ \rf{eqS1a} below), CDT turns out to have a remarkably
rich phase diagram, as function of the bare inverse gravitational coupling $\kappa_0$ and the asymmetry parameter $\Delta$. 
The existence of three distinct phases with corresponding transition lines between them is one of
the classic CDT results \cite{CDT1,CDTHL}. There are two phases $A$ and $B$ in which no meaningful (from the point of
view of General Relativity) semiclassical limit
seems to exist, a conclusion one arrives at by monitoring the dynamics of the total spatial volume of the universe in time. 
By contrast, phase $C$ does display physically interesting behaviour, in that the dynamics
generates a quantum universe whose large-scale properties match those of a four-dimensional de Sitter space. 
While the $A$-$C$ phase transition was subsequently shown to be first order, the $B$-$C$ transition turns out to be 
a second-order transition \cite{samo}, opening the exciting possibility of finding a UV fixed point and an associated
continuum theory. 
\begin{figure}[t]
\centering
\scalebox{0.85}{\includegraphics{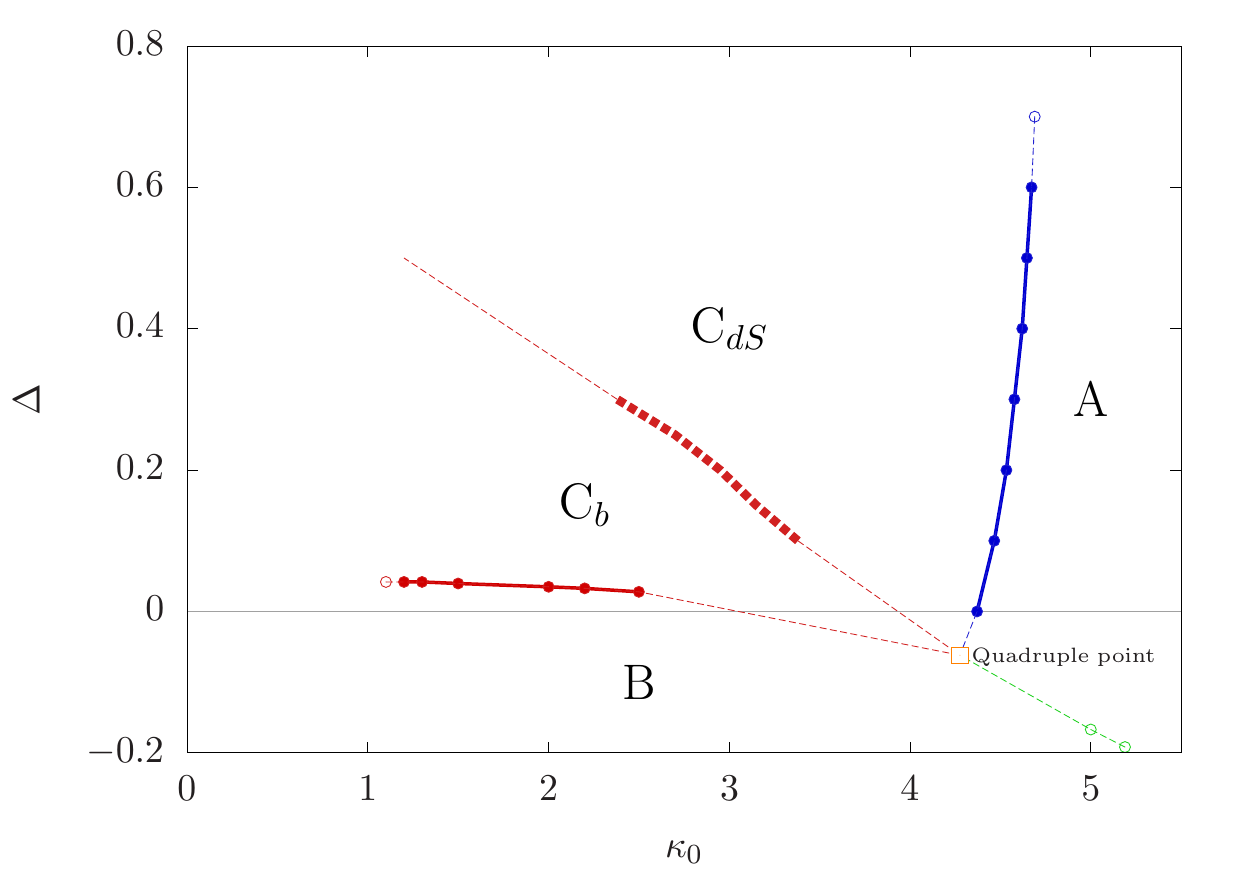}}
\caption{CDT phase diagram in terms of the bare couplings $\kappa_0$ and $\Delta$, with the phases $A$, $B$, 
the de Sitter phase $C_{dS}$ and the bifurcation
phase $C_{b}$. 
The last two and the new phase transition line separating them reflect our new, refined understanding
of CDT's phase structure. (Fat dots and squares refer to actual measurements. The ``quadruple point" is
based on extrapolation only.)
}
\label{fig1}
\end{figure}

Recently, this picture has been further refined with the discovery of a new transition line cutting diagonally through 
phase $C$ and dividing it into two regions \cite{bifurcation1,bifurcation2}, see Fig.\ \ref{fig1}. A first investigation of the
order of the new phase transition has not yielded a conclusive answer on whether it is of first or higher order \cite{cgj}.
Since it has now become clear that there are two phases instead of the single phase $C$, it is a good time to settle on a
definite name and notation for them. To ensure continuity with the previous situation and at the same time be descriptive 
we suggest ``de Sitter phase" ($C_{dS}$) for the phase above the new phase transition (``above" in the usual
$\kappa_0$-$\Delta$ phase diagram), and ``bifurcation phase" ($C_{b}$) for the phase below the transition.
The transition formerly known as the $A$-$C$ transition then becomes the $A$-$C_{dS}$ transition, and
the former $B$-$C$ transition becomes the $B$-$C_{b}$ transition. New is the de Sitter-bifurcation transition
$C_{dS}$-$C_{b}$. 

The properties of the de Sitter phase $C_{dS}$ coincide with those previously associated with phase $C$, including the
de Sitter-like scaling of the spatial volume. A de Sitter-like scaling is also observed in the bifurcation phase $C_{b}$, but is modulated
there by other dynamical effects, as became apparent when studying the behaviour of the spatial volume in the context of 
the so-called effective transfer matrix introduced in \cite{TMfirst}. In this setting one studies the CDT system with 
a minimal total number of time steps $t_{tot}$, typically $t_{tot}=2$, compared to the usual $t_{tot}=80$. While
in the latter simulations inside phase $C$ the entire (de Sitter) universe is visible, in the transfer matrix setting one only 
has access to a thin ``slice" of the universe. Of course, one has to investigate carefully to what extent both systems
describe the same physics (including phase structure and phase transitions), and to isolate finite-size and finite-time effects. 
Several of the results presented below contribute to this issue. 

A major new result found in the transfer matrix approach is the new phase transition $C_{dS}$-$C_{b}$, between a phase
where the three-volume of adjacent constant-time slices tends to align ($C_{dS}$) and a phase where the volume profile
is modulated such that the volumes of alternating slices align ($C_{b}$). The latter results in a two-peak structure when one plots the
volume-volume correlator of neighbouring slices as a function of their (oriented) volume difference \cite{bifurcation1}.
This motivated the term ``bifurcation phase", since the corresponding plot in the de Sitter phase $C_{dS}$ has only a single peak.
Below, we will uncover a dynamical mechanism behind the bifurcation transition $C_{dS}$-$C_{b}$ and give
it a more direct interpretation in terms of symmetry breaking. At the same time, this will shed some light on
the geometric nature of the bifurcation phase, which at this stage is only incompletely understood.  

The reason why such an understanding is not straightforward has to do with the nonperturbative character of the
dynamics, which is determined by the interplay between the action and the entropy, that is, the number of 
configurations (triangulated spacetimes) for given values of the action. An example of this is the behaviour of
CDT near the second-order $B$-$C_{b}$ transition. The original investigation \cite{samo} exhibited unusual
features, some of them more reminiscent of a first-order transition. Interestingly, as we will see, these first-order 
aspects disappear when one employs a different prescription for fixing the overall spacetime volume. By performing a quantitative
analysis of the entropy factor near the transition, we will give a common explanation for both of these phenomena below.

All results presented in this work contribute to the understanding of the dynamical mechanisms determining the behaviour
and phase structure of nonperturbative systems of higher-dimensional (in this case four-dimensional) geometry, about which 
relatively little is known, compared to the well-studied case of two-dimensional gravity of either signature. To the extent these
properties are driven by ``entropic effects", one would expect them to be largely independent of the details of the CDT set-up,
and therefore not necessarily confined to this particular approach to nonperturbative quantum gravity.
 
The remainder of this paper is organized as follows. After a short summary of some vital ingredients of the CDT approach
in Sec.\ \ref{nutshell}, we concentrate in Sec.\ \ref{puzzle} on the second-order $B$-$C_{b}$ phase transition. We explain a 
curious dependence of the transition signal on the choice of volume-fixing found in previous work by carefully analyzing
the entropy factor underlying this behaviour. In the appendix we show that a simple ansatz for this factor can reproduce
the characteristic shapes of the transition signals. 
Sec.\ \ref{bif} is dedicated to a closer examination of the new
bifurcation phase $C_{b}$. It is performed by simulating an ensemble of CDT configurations with minimal time extension
$t_{tot}\! =\! 2$, which is found to display the same phase characteristics and phase transitions as the more customary 
large-time ensemble. We obtain a quantitative understanding of the properties of the bifurcation phase in terms of
a vertex of very high order that appears on one of the two spatial slices of the system. This enables us to
give a direct interpretation of the $C_{dS}$-$C_{b}$ phase transition in terms of symmetry breaking, in this case,
the breaking of the 
homogeneity and isotropy of the average geometry observed in the neighbouring de Sitter phase $C_{dS}$.
A summary and conclusions are presented in Sec.\ \ref{final}.

\section{CDT set-up in a nutshell}\label{nutshell}

We will briefly review the ingredients of the CDT construction and their notation, to the extent they are needed in 
the rest of the paper. A comprehensive description of the set-up can be found in \cite{physrep}. 
The regularized CDT implementation of the path integral for pure gravity takes the form of a sum over
distinct causal triangulations $T$. After Wick rotation, it is schematically given as the partition function
\begin{equation}
Z=\sum_{T\in {\cal T}} \frac{1}{C_T}\ {\rm e}^{-S^{EH}(T)},
\label{partition}
\end{equation}
where $S^{EH}(T)$ is the Einstein-Hilbert action of the piecewise flat manifold $T$ (originally due to Regge) and 
$C_T$ denotes the order of the automorphism group of $T$, a number equal to 1 in the generic
case that the triangulation $T$ does not possess any such symmetries. 
A triangulation can be thought of as assembled from elementary building blocks, the four-dimensional
simplices, which in the standard CDT formulation come in two types, depending on their edge length
assignments. 

Recall that the interior, flat geometry of a $d$-dimensional simplex (a ``$d$-simplex") is completely fixed by
its edge lengths. CDT configurations have two types of edges, space-like and
time-like. All space-like edges have the same proper length squared $a^2$, and all time-like
edges the same proper length squared $-\alpha a^2$, where $\alpha >0$ and $a$ denotes a UV cutoff 
that will be taken to zero as the regularization is removed. After Wick-rotating, which amounts to
an analytic continuation of the parameter $\alpha$ to the negative real half-axis in the complex 
$\alpha$-plane \cite{physrep},
the triangulations still have two different edge lengths (unless $\alpha$ is set to unity), namely, 
\beql{ja2}
\ell^2_{\rm space-like} = a^2,\quad\quad  \ell^2_{\rm time-like} =\alpha \,a^2,
\eeq
where $\alpha >7/12$ to satisfy triangle inequalities.
\begin{figure}[t]
\centering
\scalebox{0.45}{\includegraphics{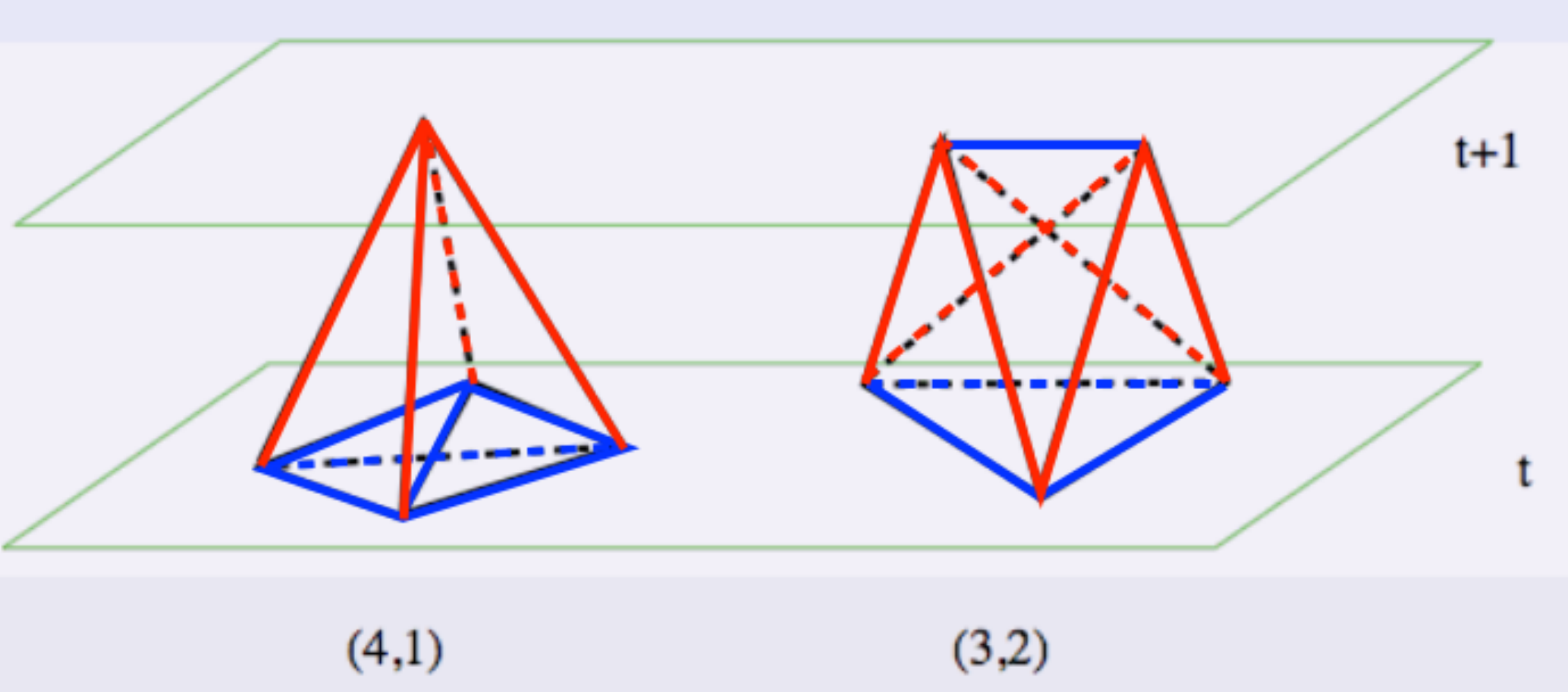}}
\caption{The two types of four-simplex appearing in CDT, the (4,1)-simplex (left) and the (3,2)-simplex
(right), interpolating between neighbouring spatial slices of constant integer time $t$. 
Space-like edges are drawn in blue, time-like ones in red.
}
\label{4simplices}
\end{figure}

In addition to the Minkowskian geometry of its simplicial building blocks, the causal character of CDT quantum
gravity is reflected in the gluing rules for the four-simplices, which are such that the causal
(= light cone) structure of each triangulation $T$ is well-defined. In standard CDT this is achieved
through the presence of a stacked structure associated with the presence of a discrete time
parameter $t$.\footnote{There is an alternative version of CDT, using so-called locally causal dynamical 
triangulations (LCDT) \cite{jordanthesis}, where the causal structure is only implemented locally, without referring
to a preferred global lattice time slicing. This can be achieved by introducing new types of building blocks (with edge
lengths still given by eq.\ (\ref{ja2})). In three spacetime dimensions, this approach has produced results compatible
with those of CDT \cite{jordanloll}, at the expense of considerable additional computational complexity.}
A causal triangulation consists of a sequence of three-dimensional spatial
triangulations, each labelled by an integer $t$, with four-dimensional space-time simplices
interpolating between adjacent slices of constant times $t$ and $t+1$. In the present work, the spatial slices
will have the topology of the three-sphere. 

The two four-simplex types mentioned
above are precisely those that are compatible with this stacked or layered structure. They are the (4,1)-simplex (together
with its time-reflection, the (1,4)-simplex) and the (3,2)-simplex (together with the time-reflected (2,3)-simplex).
A (4,1)-simplex shares a purely space-like three-simplex (spanned by four vertices) with the three-dimensional triangulation 
at time $t$ and a single vertex with the spatial triangulation at time $t+1$, whereas a (3,2)-simplex shares
a two-dimensional space-like triangle (spanned by three vertices) with the slice at time $t$ and a space-like edge
(spanned by two vertices) with the slice at time $t+1$. It follows that a (4,1)-simplex has 6 space-like and 4 time-like links, and
a (3,2)-simplex has 4 space-like and 6 time-like links (see Fig.\ \ref{4simplices}). Analogous statements hold for the (1,4)- and (2,3)-simplices
when interchanging $t$ and $t+1$. 

Since there are only two geometrically distinct building blocks, the Einstein-Hilbert-Regge action (including a
cosmological constant term) assumes a simple form in terms of the global ``counting variables" $N_i(T)$, $i=0,1, ..., 4$,
which for a given triangulation $T$ count the number of $i$-dimensional simplices contained in $T$. Below, we will
use the numbers $N_0$ of vertices and $N_4$ of four-simplices. It will be essential to keep track of
the separate numbers $N_4 ^{(4,1)}$ and $N_4^{(3,2)}$ of the two different types, where by definition these
numbers count building blocks of either time orientation, for example, $N_4 ^{(4,1)}$ is the number of (4,1)- and (1,4)-simplices
together. Since they
occur frequently in our formulas, we will use $N_{41}\! :=\! N_4 ^{(4,1)}$ and $N_{32}\! :=\! N_4^{(3,2)}$
as a shorthand notation. Of course, we have $N_{41}(T)+N_{32}(T)=N_4(T)$ for any $T$.
In terms of these, we can finally write the gravitational action as \cite{physrep}
\begin{equation}
S^{EH}(T)	 = - (\kappa_0 +6\Delta) N_0 + \kappa_4 (N_{41}+N_{32}) +\Delta (2 N_{41}+N_{32}),
\label{eqS1a}
\end{equation}
where $\kappa_0$ is the bare inverse Newton constant, $\kappa_4$ (up to a $\kappa_0$-dependent shift)
the bare cosmological constant, and $\Delta$ is an asymmetry parameter that depends
on the finite, relative scaling $\alpha$ between time- and space-like links introduced in (\ref{ja2}). Details of
this algebraic dependence will not concern us here, other than the fact that $\Delta$
vanishes for equilateral simplices, that is, $\Delta (\alpha\! =\! 1)=0$. In the nonperturbative regime investigated
by CDT, $\Delta$ plays the role of a coupling constant. To emphasize various aspects of the action
(\ref{eqS1a}), whose motivation will become clear in subsequent sections, we can rewrite it
in a number of equivalent ways,
\begin{align}
S^{EH}(T)	& \equiv - \kappa_0  N_0 + (\kappa_4 +\Delta) N_4 +\Delta (N_{41}- 6 N_0)
\label{eqS1s}\\
& \equiv - (\kappa_0 +6\Delta) N_0 + \bigg( \kappa_4+\frac{3\Delta}{2}\bigg) N_4 +\frac{\Delta}{2}\, x
\label{eqS1b}\\
&\equiv  - \kappa'_0 N_0 + {\kappa}_{41} N_{41} +\kappa_{32} N_{32}.
\label{eqS1c}
\end{align}
Eq.\ (\ref{eqS1s}) is a straightforward reshuffling of terms, eq.\ (\ref{eqS1b}) is a
rewriting of (\ref{eqS1a}) using the difference $x:=N_{41}-N_{32}$,
while (\ref{eqS1c}) results after performing a linear redefinition of the coupling constants according to
$\kappa'_0:=\kappa_0+6\Delta$, $\kappa_{41}:=\kappa_4+2\Delta$ and $\kappa_{32}:= \kappa_4+\Delta$.

In the actual CDT computer simulations the lattice volume is kept (approximately) constant, by adding a 
volume-fixing term $S_{\it fix}$ to the bare action \rf{eqS1a}. This means there are de facto only two tunable bare couplings,
$\kappa_0$ and $\Delta$, as illustrated by the phase diagram of Fig.\ \ref{fig1}. Two different quadratic
volume fixings have been used in the literature, either fixing the total number of four-simplices to $\bar N_4$ by
setting 
\begin{equation}
S_{\it fix}^{\bar N_4}(N_4)=\varepsilon (N_4-\bar N_4)^2
\label{sfix1}
\end{equation}
or fixing the number of (4,1)-simplices to some target value $\bar N_{41}$ by setting
\begin{equation}
S_{\it fix}^{\bar N_{41}}(N_{41})=\varepsilon (N_{41}-\bar N_{41})^2,
\label{sfix2}
\end{equation}
where $\varepsilon$ in both cases denotes an appropriately chosen small, positive parameter. 
Inside the ``old" phase $C$ and well away from the phase
transitions $B$-$C_{b}$ and $A$-$C_{dS}$ one does not expect results to depend on the
volume fixing used, since at a given $(\kappa_0,\Delta)$ the two four-simplex types
occur approximately in a fixed ratio \cite{physrep}. However, as already mentioned above, some
measurements at the $B$-$C_{b}$ transition appear to depend on the volume fixing, a phenomenon
that will be explained in the following Sec.\ \ref{puzzle}.

\section{A second look at the $B$-$C_{b}$ transition}
\label{puzzle}

We begin by examining the transition between phase $B$ and the bifurcation phase $C_{b}$.
It has been known for some time to be a second-order transition, and thus potentially interesting
for continuum physics. 
The original investigation of what was then called the $B$-$C$ transition was performed at fixed $N_4$,
implemented by a volume fixing of the form (\ref{sfix1}), for volumes of up to $N_4=160k$ \cite{samo}.
The order parameter chosen to study the transition was conj$(\Delta):=N_{41}-6 N_0$, which
is the expression conjugate to $\Delta$ at fixed $N_4$, as can be read off from (\ref{eqS1s}).
The analysis required some care, because the probability distribution of conj$(\Delta)$ measured 
at the transition exhibited a double-peak structure.
This is unusual, because a double peak is typically associated with a {\it first}-order transition, 
where it is brought about by a jumping of the
order parameter between two metastable states on either side of the transition. However,
in the case at hand 
a careful analysis of finite-size effects in terms of observables like the Binder cumulant,
particularly suited to distinguishing between first- and higher-order transitions,
all pointed towards a second-order transition. 

\begin{figure}[t]
\centering
\includegraphics[width=0.8\textwidth]{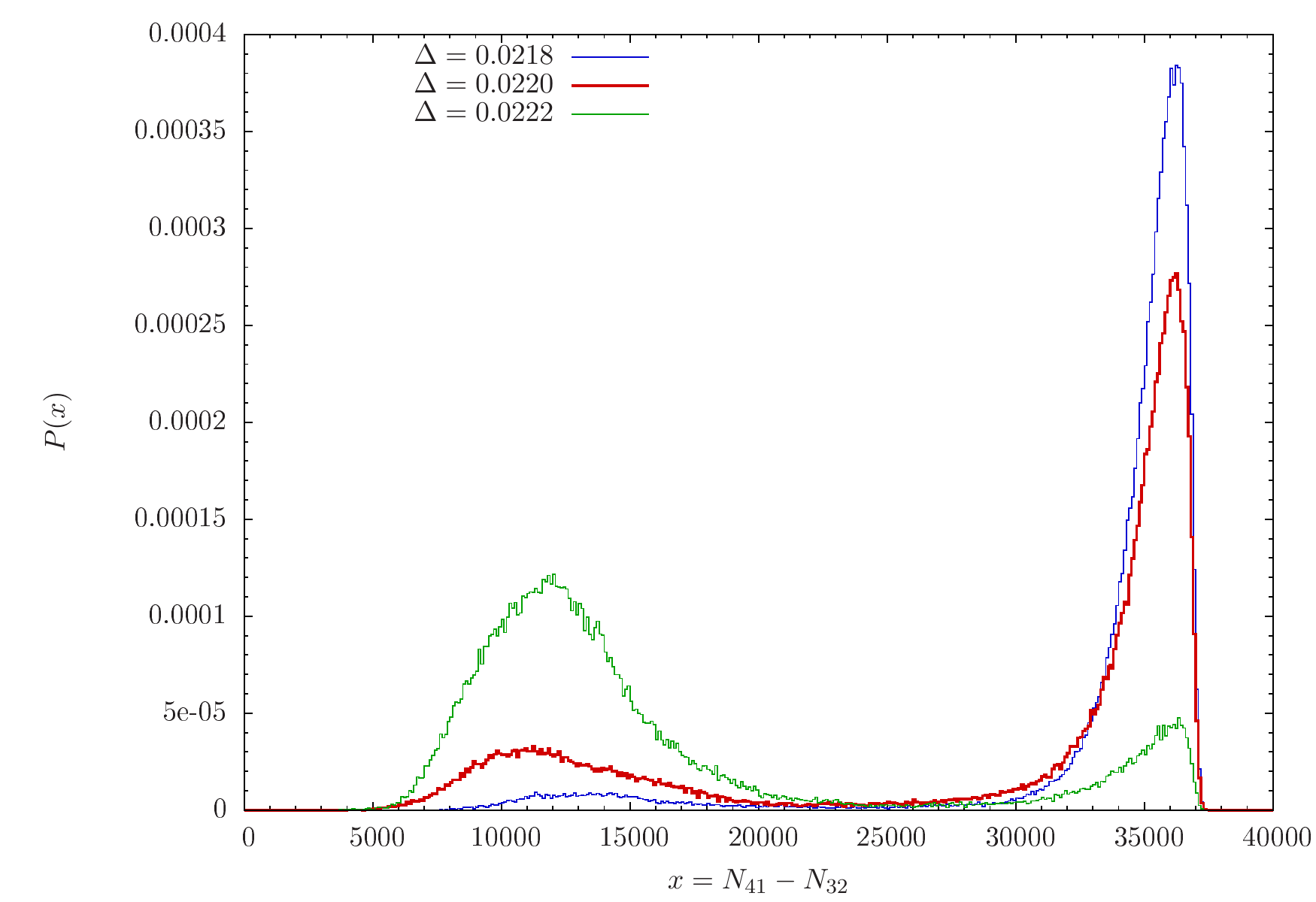}
\caption{
	Probability distribution $\bar{P}(x)$ of the order parameter $x$, measured
	at three different couplings $\Delta$ close to the critical point $\Delta_c\approx 0.0220$, for total volume $N_4 = 40k$ 
	and $\kappa_0=2.2$.}
\label{fig:distN4}
\end{figure}

We have found it convenient to work with another
order parameter, the quantity $x=N_{41}-N_{32}$ introduced earlier. Looking
at the action (\ref{eqS1b}), one observes that $x$ would be conjugate to $\Delta$ for fixed $N_4$ if we also held
$N_0$ fixed (which we do not). Using $x$ instead of conj$(\Delta)$ as an order parameter corresponds to
approaching the transition line along a slightly different phase space trajectory, and
leads to an equivalent result for its probability distribution $\bar{P}(x)$.\footnote{We will use an over-bar
notation $\bar{P}(x)$ for the distribution at fixed $N_4$ and an over-tilde notation $\tilde{P}(x)$
for the distribution at fixed $N_{41}$.} The results 
for $\bar{P}(x)$, measured at fixed $N_4=40k$ and for time extension $t_{tot}=80$, 
are shown in Fig.\ \ref{fig:distN4} and display the same kind of double peak as 
in the original work \cite{samo}. Note that the relative height of the two peaks in the distribution 
$\bar{P}(x)$ depends on the coupling $\Delta$. We define the critical value $\Delta_c$ as
the value where the peaks have the same height\footnote{Alternatively, one could define
$\Delta_c$ as the point where the areas under the two peaks become
equal. The resulting $\Delta_c$ differs only slightly from the ``equal-height 
$\Delta_c$''. }. 

Following a spacetime configuration and measuring its $x$-value as a function of Monte Carlo time, 
one finds that $x$ is located close to one of the 
peaks for some time and then makes a very rapid change to the other peak where
it again stays for some time. (Examples of Monte Carlo 
time histories of order parameters, albeit in a slightly different context, are depicted in Fig.\ \ref{nilas15} below.) 
This is precisely the behaviour expected at a first-order transition, for sufficiently small volumes. 
However, for a genuine first-order transition such a cross-over between different phases will be suppressed 
as the system size goes to infinity. The absence of such a behaviour for increasing volume
led to the more detailed investigation of \cite{samo}, with the outcome that 
the $B$-$C_{b}$ transition in CDT appears to be of higher order.
 
Somewhat surprisingly, when repeating the same measurements with $N_{41}$ rather than $N_4$ kept fixed,
we found no trace of a double peak structure for any of the order parameters considered.
The distribution of $x$ (which for constant $N_{41}$ coincides with the distribution of $N_{32}$) 
is shown in Fig.\ \ref{fig:distN41}. 
As explained in more detail in subsection \ref{sec:trann41} below, we have determined the (pseudo-)critical value $\Delta_c$
from a peak in the susceptibility $\chi(x)=\langle x^2\rangle - \langle x\rangle^2$ under variation of $\Delta$, 
where the distribution $\tilde{P}(x)$ has maximal width. 
Thus it appears that for fixed $N_{41}$ the situation is consistent with that of a typical second-order transition.
\begin{figure}
\centering
\scalebox{0.7}{\includegraphics{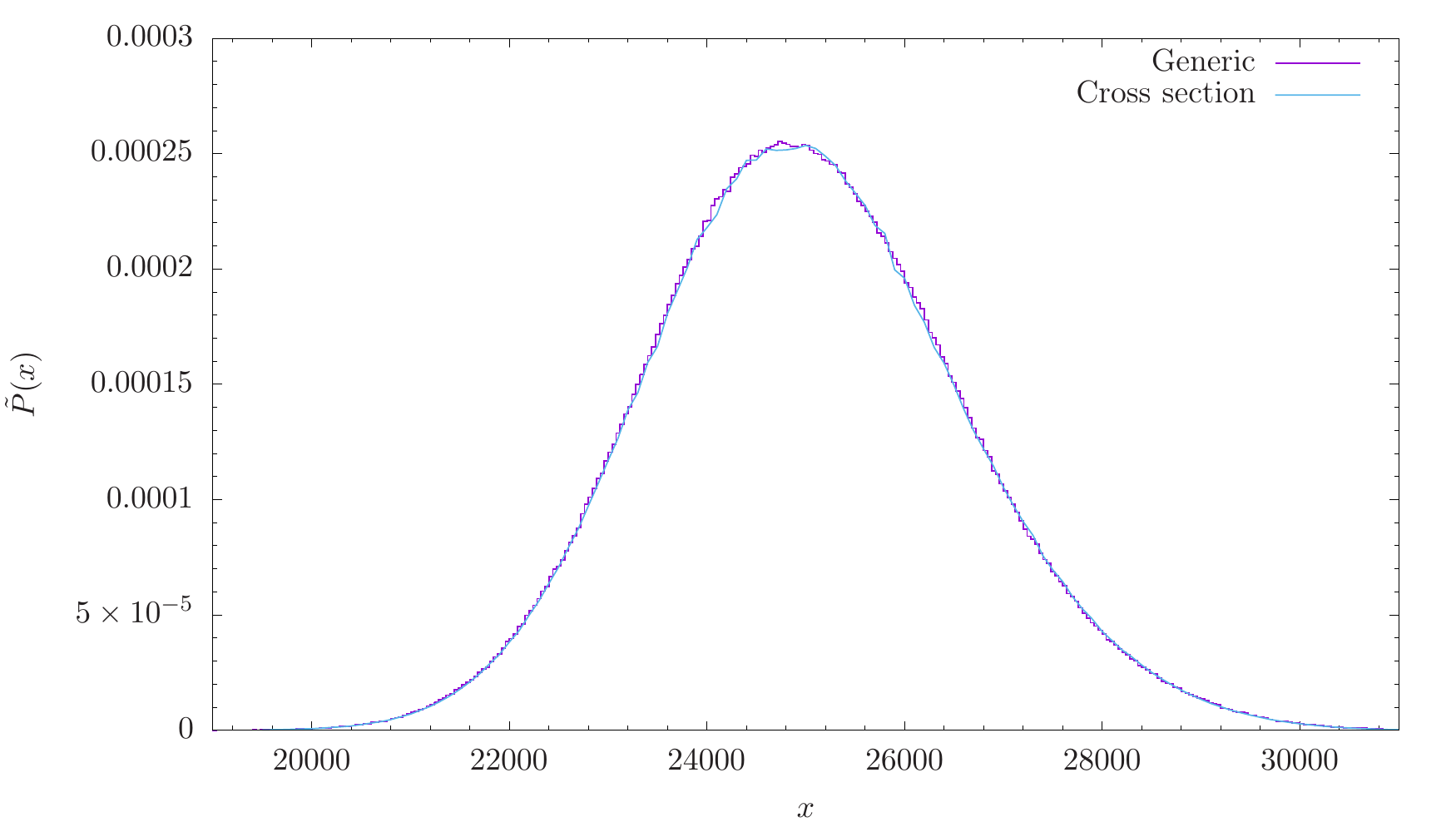}}
\caption{
	For fixed volume $N_{41}$, the probability distribution $\tilde{P}(x)$ does not have a
	double-peak structure close to the critical point $\Delta_c\approx 0.0220$.
	The violet curve shows Monte Carlo measurements taken at $N_{41}\! =\! 33k$, for $\kappa'_0\! =\! 2.3320, \kappa_{41}\! =\! 0.9856$ and
	$\kappa_{32}\!  =\! 0.9636$
	(couplings defined below eq.\ \rf{eqS1c}). The blue curve represents the cross section (\ref{jnew2}), see Sec.\ \ref{sec:trann41}.
	}
\label{fig:distN41}
\end{figure}

In what follows, we will demonstrate that the observed dependence of the distribution of $x$ on the
volume fixing has its origin in what we will call the entropy (factor) $\cN(N_0,N_{41},N_{32})$, 
the function that counts the
number of configurations (including their symmetry factors $1/C_T$) for given values
of the counting variables $N_0$, $N_{41}$ and $N_{32}$, namely,
\beql{eq:N}
	\cN(N_0, N_{41}, N_{32}) =
	\sum_{T \in \cT ( N_0,N_{41},N_{32})}  \frac{1}{C_{\cT}},
\eeq
where $\cT( N_0,N_{41},N_{32})$ denotes the set of triangulations with 
fixed $N_0$, $N_{41}$ and $N_{32}$. Using the action in the form (\ref{eqS1c}),
the partition function can now be written as
\beql{ja4}
 Z(\kappa'_0,\kappa_{41},\kappa_{32}) = 
\sum_{N_0,N_{41},N_{32}} \e^{-S(N_0,N_{41},N_{32})} \cN(N_0,N_{41},N_{32}) .
\eeq
We will apply Monte Carlo techniques to extract the entropy $\cN(N_0, N_{41}, N_{32})$.
In order to measure this function over a whole range of values in the $(N_{41}, N_{32})$-plane, 
as we would like to do, an efficient method is to modify the action in a controlled way 
such that one probes smaller regions. By adding quadratic terms 
\beql{ja10} S_{\it fix}^{\bar{N}_{41}, \bar{N}_{32}}({N}_{41}, {N}_{32}) = \varepsilon (N_{41} - \bar{N}_{41})^2 + \varepsilon (N_{32} - \bar{N}_{32})^2
\eeq
to the action \rf{eqS1c}, one ensures that the Monte Carlo simulations
probe a well-defined, not too large region in the vicinity of a prescribed 
point $(\bar{N}_{41}, \bar{N}_{32})$. More specifically,
a given set of numbers $N_0$, $N_{41}$ and $N_{32}$ will occur with probability 
\begin{align}
P_{\bar{N}_{41},\bar{N}_{32}}(N_0, N_{41}, N_{32}) &\propto \cN(N_0, N_{41}, N_{32}) \cdot 
\e^{- S(N_0, N_{41}, N_{32})  - S_{\it fix}^{\bar{N}_{41}, \bar{N}_{32}}({N}_{41}, {N}_{32})     } .
\label{eqP}
\end{align}
We have covered the region of interest by eight patches corresponding to different 
values $\bar{N}_{41}, \bar{N}_{32}$, such that they overlap mutually.
This allows us to adjust the relative probability distributions measured in the 
different patches to a common probability distribution, which is determined up to 
a common normalization factor. We could in principle have chosen different 
values for the three couplings $\kappa'_0$, $\kappa_{41}$ and $\kappa_{32}$ 
in the various patches, but we keep them constant across all patches 
and equal to the reference values 
$\bar{\kappa}'_0$, $\bar{\kappa}_{41}$ and $\bar{\kappa}_{32}$.

To simplify the comparison between fixing $N_4$ and $N_{41}$, we 
integrate out the number $N_0$ of vertices weighted by $e^{\bar{\kappa}'_0 N_0}$
to obtain the ``reference'' probability distribution
\beql{ja11}
{\cal P}(N_{41}, N_{32}) := \cC \cdot \sum_{N_0} \cN(N_0, N_{41}, N_{32}) \cdot \e^{\bar{\kappa}'_0 N_0 
- \bar{\kappa}_{41} N_{41} - \bar{\kappa}_{32} N_{32}},
\eeq
where the normalization factor $\cC$ ensures that the probabilities add up to one. The
distribution (\ref{ja11}) can be extracted from the measured quantities $P_{\bar{N}_{41},\bar{N}_{32}}(N_0, N_{41}, N_{32})$ according to
\beql{ja12} 
	{\cal P}(N_{41}, N_{32}) = \tilde{\cC} \cdot \sum_{N_0} P_{\bar{N}_{41},\bar{N}_{32}}(N_0, N_{41}, N_{32}) 
	\cdot 	\e^{S_{\it fix}^{\bar{N}_{41}, \bar{N}_{32}} ({N}_{41},{N}_{32})}.
\eeq	
It is understood that during the matching process for the overlap regions the various $P_{\bar{N}_{41},\bar{N}_{32}}$ 
have been normalized relative to each other such that after multiplication with ${\rm exp}(S_{\it fix})$ and
summing over $N_0$ only a single common normalization factor $\tilde{\cC}$ is needed, as already mentioned
above. The right-hand side of eq.\ (\ref{ja12}) therefore describes a single, joint probability distribution, which 
by construction no longer depends on ${\bar N}_{41}$ and ${\bar N}_{32}$.

\begin{figure}
\includegraphics[width=0.95\textwidth]{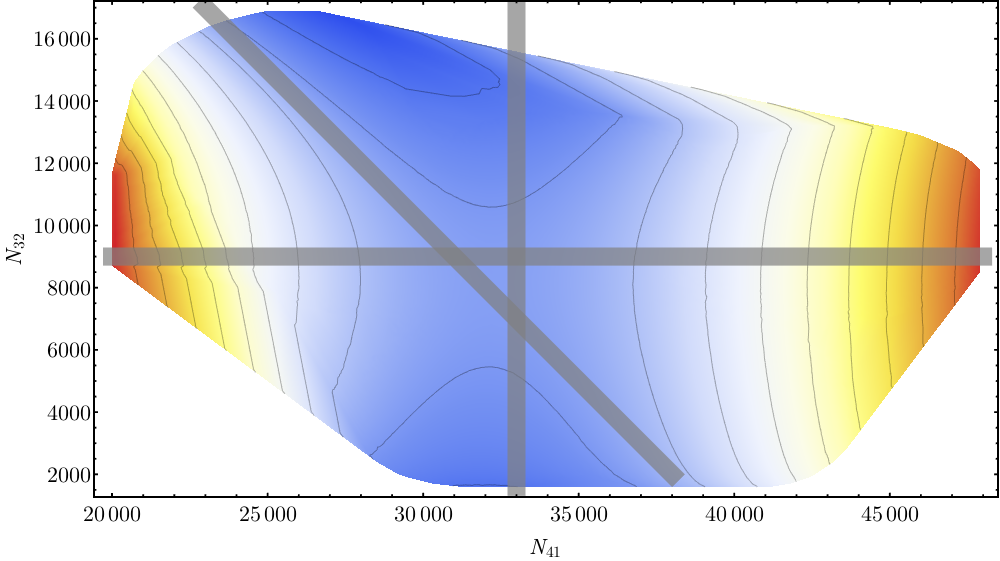}
\caption{The free energy $F(N_{41}, N_{32})$ for
	$\bar{\kappa}'_0 \! =\! 2.3320$, $\bar{\kappa}_{41} \! =\! 0.9856$ and $\bar{\kappa}_{32} \! =\! 0.9636$.
	Its value increases from blue to red.
	The grey lines represent the cross sections considered in the text.}
\label{fig:F}
\end{figure}

Rather than working directly with ${\cal P}(N_{41}, N_{32})$, we have found it 
convenient to work with its logarithm
\begin{align}
\label{eq:F}
F(N_{41}, N_{32}) &:=  \log {\cal P}(N_{41}, N_{32}) \\
&= - \bar{\kappa}_{41} N_{41} - \bar{\kappa}_{32} N_{32} + \log \sum_{N_0} \cN(N_0, N_{41}, N_{32}) \cdot e^{\bar{\kappa}'_0 N_0}, \nonumber
\end{align}
which can be interpreted as (minus) the free energy of the system.
The density plot of the measured free energy (\ref{eq:F}) as a function of $N_{41}$ and $N_{32}$ is shown in Fig.\ \ref{fig:F}. 
Simulations were performed at $\bar{\kappa}'_0 \! =\! 2.3320, \bar{\kappa}_{41}\! =\! 0.9856$ and $\bar{\kappa}_{32}\!  =\! 0.9636$,
corresponding to the critical point on the $B$-$C_{b}$ transition line
observed in simulations with fixed $N_4$ and $N_{41}$. 
The colours run from blue, corresponding to low values of the free energy $F(N_{41}, N_{32})$ and thus of the probability 
${\cal P}(N_{41}, N_{32})$, to red, indicating high values of $F$ and $\cal P$.
Note that the function $F(N_{41}, N_{32})$ has a saddle point at the centre of the region considered;
it is convex for $N_{32}\! =\! \const$ (horizontal line in Fig.\ \ref{eq:F}) and concave for $N_{41}\! =\! \const$ (vertical line).
We will show below that this shape explains the different behaviour of the probability distributions $\bar{P}(x)$ 
and $\tilde{P}(x)$ of the order
parameter $x$, depending on whether $N_4$ or $N_{41}$ is kept fixed in the simulations.

\subsection{Double-peak structure for fixed $N_4$}

\begin{figure}
\centerline{\includegraphics[width=0.70\textwidth]{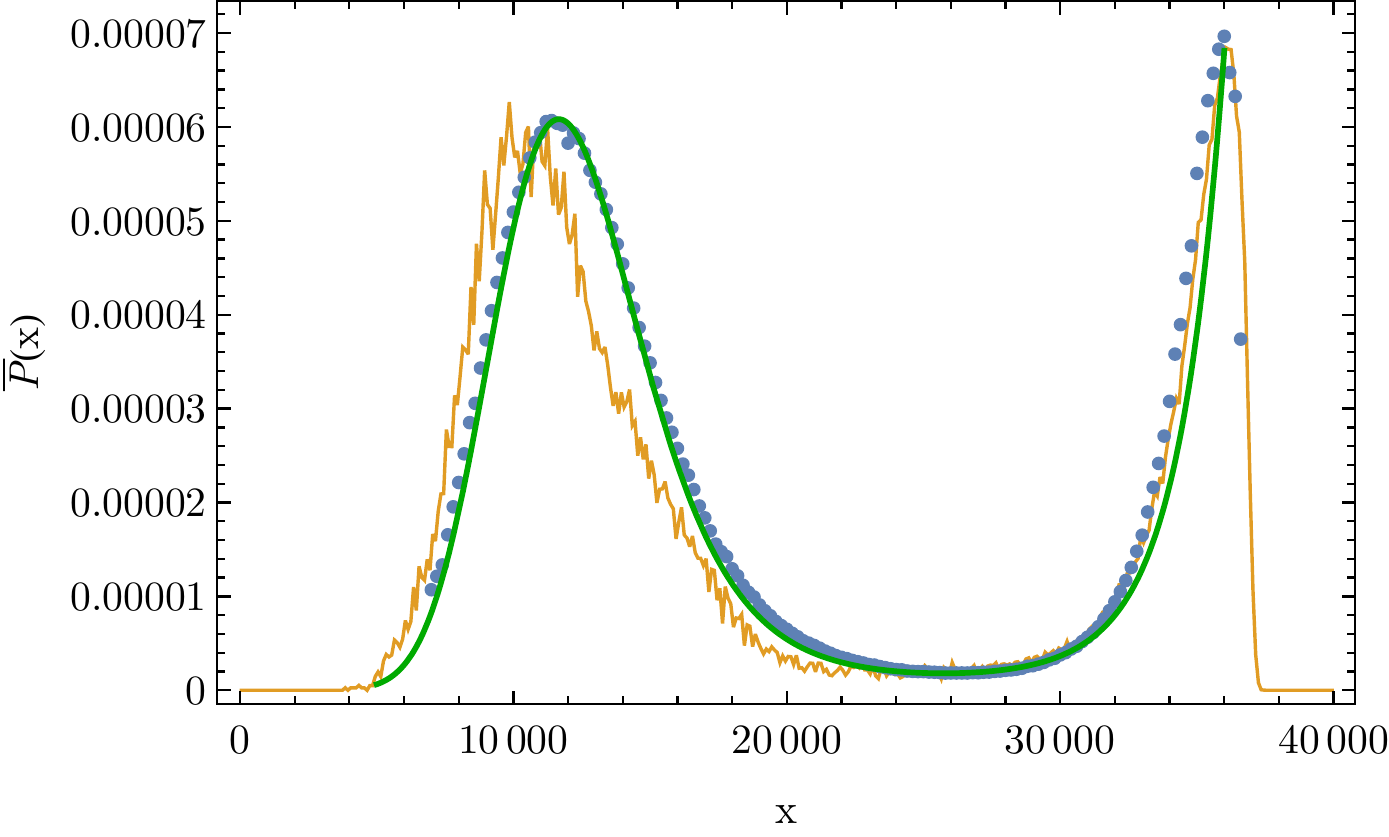}}
\caption{Distribution $\bar{P}(x)$ of 
the order parameter $x$ for fixed $N_4\! =\! 40k$: direct measurement from Monte Carlo data (yellow curve; 
$\Delta$ adjusted to obtain peaks of equal height),
calculated from the free energy $F(N_{41}, N_{32})$ according to eq.\ \rf{ja14}
	(blue dots), and obtained from a model function for the free energy (green curve), see the Appendix for further details.
	}
\label{fig:FN4}
\end{figure}

In connection with Fig.\ \ref{fig:distN4} we already reported on the double peak in direct Monte
Carlo simulations of the probability distribution $\bar{P}(x)$ observed for fixed $N_4$. Remarkably,  
the same double peak can be reproduced by taking a cross section along the diagonal grey line 
$N_4\! =\! 40k$ indicated in Fig.\ \ref{fig:F}, and extracting a probability distribution $\bar{P}(x)$ 
from the measured values $F(N_{41}, N_{32})$ according to 
\beql{ja14}
\bar{P}(x) = {\cal P}\left(N_{41}\! =\! \frac{N_4 + x}{2}, N_{32}\! =\! 
\frac{N_4 - x}{2} \right)  =  {\rm exp}\left(F \left(\frac{N_4 + x}{2}, \frac{N_4 - x}{2}   \right) \right),
\eeq
where again $x=N_{41}-N_{32}$.
This is illustrated in Fig.\ \ref{fig:FN4} (blue dotted curve). 
The fact that we can reconstruct the double peak in this way shows that 
the saddle-shaped geometry of the free energy $F(N_{41}, N_{32})$ is responsible for 
this structure. In other words, in the volume range considered, the occurrence of such a double peak
is caused by ``entropy", in the sense of the distribution of configurations contributing to the path integral,
and is not an indication of the presence of a first-order transition.

\subsection{Single-peak structure and transition for fixed $N_{41}$}
\label{sec:trann41}

By contrast, for fixed $N_{41}$, implemented by adding the volume-fixing term (\ref{sfix2}) to the action, the distribution $\tilde{P}(x)$ is 
well approximated by a concave function with a single ``Gaussian-like'' bump
as illustrated by Fig.\ \ref{fig:distN41}.
The violet curve shows the results of standard Monte Carlo simulations for $\tilde{P}(x)$, while the blue line represents  
\beql{jnew2}
\tilde{P} (x) = {\cal P}(N_{41}\! =\! \bar{N}_{41},N_{32}\!=\!\bar{N}_{41}-x ) =  e^{F(\bar N_{41}, \bar{N}_{41}-x)}, \quad 
\bar{N}_{41} = 33k.
\eeq
The corresponding cross section through the $({N}_{41},N_{32})$-plane is given by the vertical grey 
line in Fig.\ \ref{fig:F}. The two methods for determining this distribution are in perfect agreement.
Note also that the maximum of $\tilde{P}(x)$ of Fig.\ \ref{fig:distN41}
and the minimum of $\bar{P}(x)$ of Fig.\ \ref{fig:distN4} occur approximately at the same point, namely,  
$N_{41} = 33k,\ N_{32} = 8k$.
\begin{figure}
\centering
\scalebox{0.8}{\includegraphics{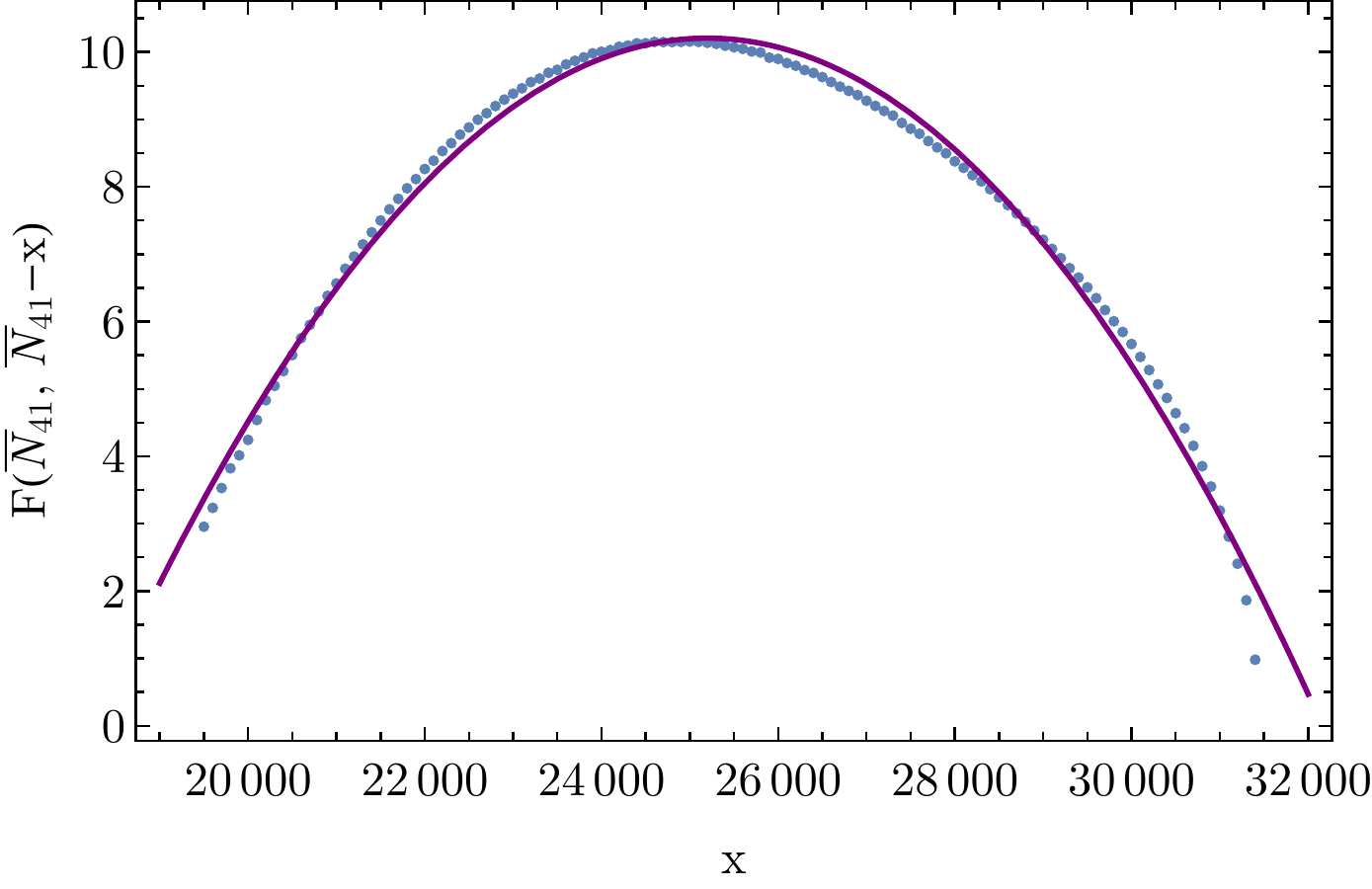}}
\caption{Measurement of the free energy $F(\bar{N}_{41}, \bar{N}_{41}-x )$ along a line of constant $\bar{N}_{41} = 33k$ (blue dots),
together with a quadratic best fit (continuous curve).}
\label{fig:vertical}
\end{figure}

\begin{figure}
\centering
\scalebox{0.8}{
\includegraphics{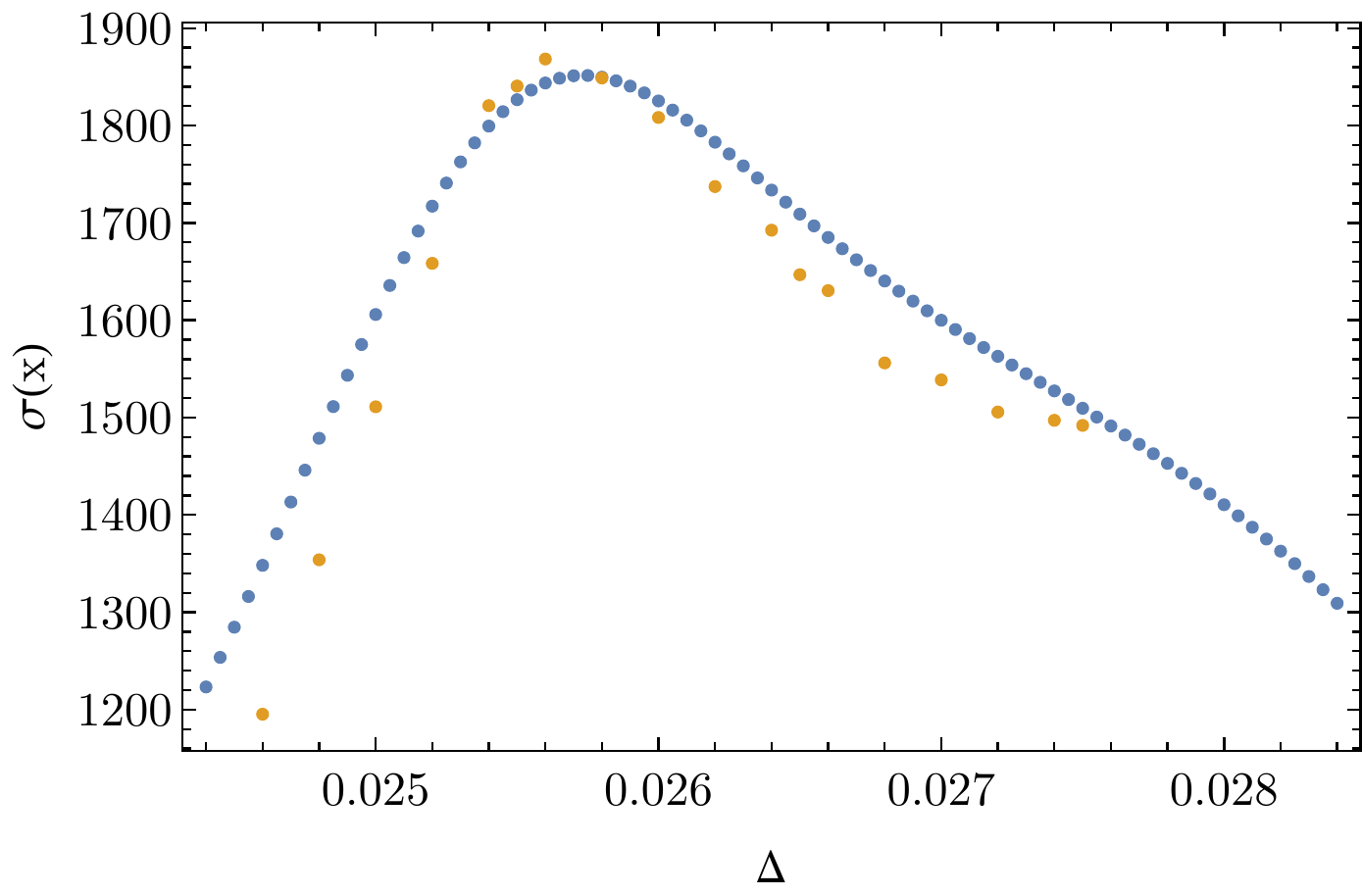}}
\caption{
	Standard deviation $\sigma (x)$ of the distribution $\tilde{P}(x)$ 
	as a function of $\Delta$ extracted from actual Monte Carlo 
	simulations (yellow sparse dots) by varying $\Delta$, as well as from $F(N_{41}, N_{32})$ 
	(blue dense dots) for constant $N_{41}$. 
}
\label{fig:deviation}
\end{figure}

The free energy $F(\bar{N}_{41}, \bar{N}_{41}-x)$, together with a quadratic fit, is shown in Fig.\ \ref{fig:vertical}. 
As mentioned earlier, by looking at where the standard deviation $\sigma (x)$ of the distribution $\tilde{P}(x)$ for $N_{41}\! =\! \bar{N}_{41}$ 
peaks as function of the coupling $\Delta$, we can extract the critical value of $\Delta$.  
To obtain the standard deviation of $\tilde{P}(x)$ one can proceed in two different ways.
One option is to simply perform Monte Carlo simulations at fixed $N_{41}$ for a number of selected values of 
$\Delta$ (yellow dots in Fig.\ \ref{fig:deviation}). The other procedure (whose results are represented by the blue dots) is more indirect and involves
a reconstruction from measurement data taken at {\it fixed} $\Delta$. 

More specifically, we have taken as a starting point
the distribution $\tilde{P}(x)$ displayed in Fig.\ \ref{fig:distN41}, which was measured for fixed 
$\bar{\kappa}_0' = 2.3320$, $\bar{\kappa}_{41} = 0.9856$ and 
$\bar{\kappa}_{32} = 0.9636$, and therefore corresponds to the single, fixed value $\bar \Delta := \bar{\kappa}_{41}\! -\! \bar{\kappa}_{32} = 0.0220$.
Since $N_{41}$ is kept fixed, the relevant coupling constants are $\bar{\kappa}_0'$ and $\bar{\kappa}_{32}$.
Due to the simple form of the action \rf{eqS1c}, there is an easy relation 
which allows us to construct from the 
distribution $\tilde P_{\bar{\kappa}_{32}} (x)$
at some fixed $\bar{\kappa}_{32}$ the distribution $\tilde P_{\kappa_{32}}(x)$ of any other value $\kappa_{32}$ (while leaving
$\bar{\kappa}_0'$ and $\bar{\kappa}_{41}$ unchanged), namely, 
\beql{ju1}
\tilde P_{\kappa_{32}}(x) \propto \tilde P_{\bar{\kappa}_{32}} (x) \; \e^{(\bar{\kappa}_{32} -\kappa_{32}) (\bar{N}_{41}-x)}.
\eeq
Since we are keeping $\kappa_{41}$ fixed, a change in $\kappa_{32}$ is equivalent to a change in $\Delta$,
in the sense that $\Delta =\bar \Delta + \bar{\kappa}_{32}-\kappa_{32}$,
which is exactly what we are interested in when determining the standard deviation $\sigma (x)$ of $\tilde{P}(x)$.

The only limitation to be taken into account when constructing $\sigma(x)$ from
numerical data in this way is that $\kappa_{32}$ should not differ too much from $\bar{\kappa}_{32}$.
One typically has accurate measurements of $\tilde P_{\bar{\kappa}_{32}} (x)$ only for some limited range in $x$,
which means that for $| \bar{\kappa}_{32}\! -\!\kappa_{32}|$ too large the centre of $\tilde P_{\kappa_{32}}(x)$ 
will be shifted to an $x$-interval where $\tilde P_{\bar{\kappa}_{32}}(x)$ is poorly determined, and thus will
lead to a large uncertainty in the derived distribution $\tilde P_{\kappa_{32}}(x)$.
As can be seen in Fig.\ \ref{fig:deviation}, in the case at hand the two very different ways of determining the standard deviation
agree remarkably well, especially with regard to the location of their peaks. This has allowed us to extract the critical
value of $\Delta$ with good accuracy as $\Delta_c \!\approx\! 0.026$. The fact that this differs slightly from the measurement
at fixed $N_4$ is not particularly surprising, since at finite volume the two volume-fixings lead to systems with
different behaviour.

In the appendix, we make a simple ansatz for the free energy $F(N_{41},N_{32})$ in terms of several free functions at most quadratic
in $N_{41}$ and $N_{32}$, which we determine uniquely from fitting them to our data. This ansatz reproduces the features 
described in this section: a cross section $N_4\! =\! const$ results in a double-peak structure and a cross section $N_{41}\! =\! const$ 
in a single-peak structure for the probability distribution of $x=N_{41}-N_{32}$. At the same time, the ansatz is too simple to reproduce 
the observed higher-order critical behaviour at the transition. This demonstrates explicitly that the unusual double-peak structure near 
the $B$-$C_{b}$ transition is not necessarily related to any critical behaviour and the question whether the observed transition is of first
or second order.

\section{The bifurcation phase}
\label{bif}

Having exhibited one aspect of the nonperturbative dynamics of CDT near the $B$-$C_{b}$ transition, we now turn to a
closer analysis of the bifurcation phase $C_{b}$, including the associated, new $C_{dS}$-$C_{b}$ transition. 
The results we will discuss are obtained in the framework of the so-called effective transfer matrix \cite{TMfirst},
which was instrumental in the discovery of the bifurcation phase in the first place \cite{bifurcation1}.
This formulation involves the reduced transfer matrix $M$, whose matrix elements 
\begin{equation}
\langle m|M|n\rangle ,\;\;\; m= N_3(t),\; n=N_3(t+1),
\label{trama}
\end{equation}
describe the transition amplitudes between a spatial configuration of three-volume $m$ at time $t$ and a neighbouring
spatial configuration of three-volume $n$ at time $t+1$. They are obtained by measuring the probabilities
\begin{equation}
P^{(2)}(m,n):=\frac{ \langle m|M|n\rangle\langle n|M|m\rangle }{{\rm Tr}\, M^2}
\end{equation}
for a system with a total time extension $t_{tot}=2$ \cite{bifurcation1} and extracting the matrix elements
according to
\begin{equation}
\langle m|M|n\rangle \propto \sqrt{P^{(2)}(m,n)}.
\end{equation}
The term {\it reduced} or {\it effective} transfer matrix refers to the fact that of all the geometric degrees of freedom that
characterize the three-dimensional spatial slices of constant integer time, 
one only keeps track of the total three-volume $N_3(t)$
of the slices at constant $t$. It is a nontrivial finding that one can reconstruct the well-known effective, ``minisuperspace" action and the global dynamics
of the three-volume \cite{CDT1,CDT2} from measurements of the reduced transfer matrix alone \cite{TMfirst,bifurcation1}. 
It was a closer examination of the ``unphysical" phases $A$ 
and, more specifically, $B$ in terms of the effective transfer matrix and the associated effective actions
that led to the discovery of the new bifurcation phase \cite{bifurcation1}.

We will study this new phase by concentrating on the correlations between 
neighbouring spatial slices. To facilitate the investigation and allow for large spatial 
slices we will consider the situation $t_{tot}\! =\! 2$ with just two spatial slices and periodic boundary conditions.\footnote{More precisely,
we work with $t_{tot}\! =\! 4$, where the spacetime geometry between $t\! =\! 2$ and $t\! =\! 4$ is an identical copy
of the geometry between $t\! =\! 0$ and $t\! =\! 2$. This is done
to maintain a regular triangulation, where by definition any (sub-)simplex is uniquely identified 
by its vertices, and happens purely for the convenience of our computer code.} Furthermore, we will keep $N_4$ fixed by including
a term (\ref{sfix1}) in the action, and set $\kappa_0\! =\! 2.2$ throughout.

\begin{figure}
\centering
\scalebox{0.28}{\includegraphics{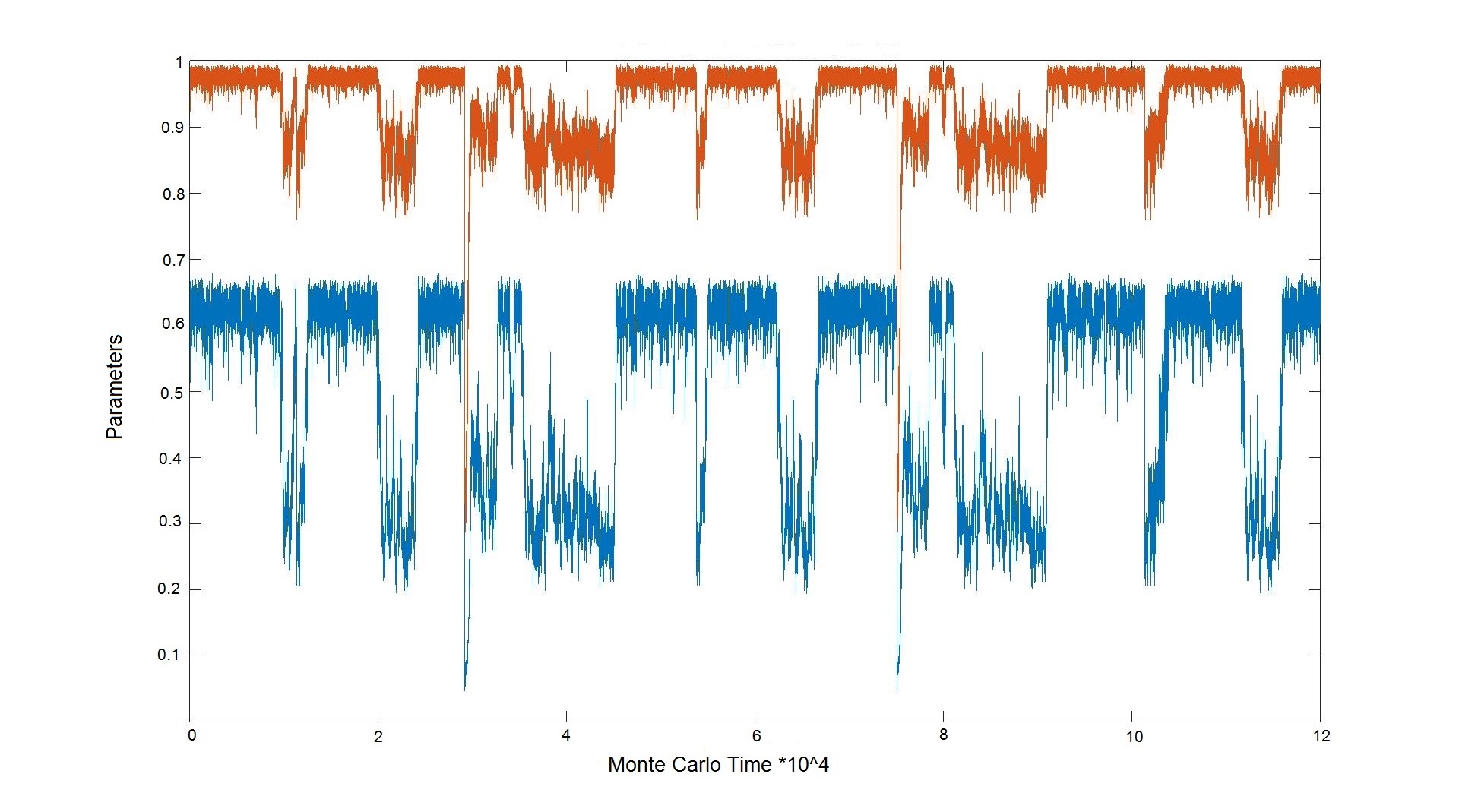}}
\caption{Two order parameters measured at the $B$-$C_{b}$ transition, at $N_4\! =\!10k$, minimal time extension $t_{tot}\! =\! 2$
and  $\kappa_0=2.2$:
order of the highest-order vertex, normalized to lie between 0 and 1 (upper graph), and conj$(\Delta )/N_4$, 
the variable conjugate to $\Delta$ (lower graph), both plotted as functions of Monte Carlo time.}
\label{nilas15}
\end{figure}

\subsection{Equivalence with large-time simulation}

Note that imposing periodic boundary conditions in time can be viewed formally as studying the system at a finite 
temperature that is inversely proportional to the time period $t_{tot}$. 
Certain phase transitions may disappear when the temperature increases and the time period therefore decreases.
However, in previous computer simulations for $t_{tot}\! =\! 4,\, 6$ we found no indications that the presence of the $B$-$C_{b}$ transition depends
on $t_{tot}$ \cite{TMfirst}. Also for the minimal time extension $t_{tot}\! =\! 2$ used here we still see a clear transition signal.
By way of illustration, Fig.\ \ref{nilas15} shows the measurements of two different order parameters at the $B$-$C_{b}$ transition, 
for $N_4\! =\! 10k$ kept fixed. One of them is the order of the highest-order vertex in the triangulation $T$, where ``order" is defined here as 
the number of one-dimensional edges sharing the vertex, normalized to lie between 0 and 1.\footnote{Many different definitions of
``vertex order" and normalization are possible, leading to qualitatively similar results. The normalization chosen here is a division by
the maximal number of edges that could meet at a vertex in a triangulation that has the same numbers of vertices in 
the two spatial slices as the given triangulation $T$. This theoretical maximum would entail that the vertex is connected by an edge to every 
other vertex in the same spatial slice and to every vertex in the neighbouring spatial slice.}
The other one is a normalized version of the quantity conj$(\Delta)\! :=\! N_{41}\! -\! 6 N_0$ introduced at the beginning of
Sec.\ \ref{puzzle}.  
As also discussed in Sec.\ \ref{puzzle}, at fixed $N_4$ and {\it large} $t_{tot}$ one finds 
a double-peak structure in the probability distribution of the order parameter $x\! =\! N_{41}\! -\! N_{32}$, 
superficially resembling the behaviour encountered at a first-order transition. 
Our observations for {\it small} $t_{tot}$ are entirely compatible with this picture, in the sense that the order
parameters depicted in Fig.\ \ref{nilas15} also display a typical first-order behaviour, jumping back and forth between
two different states on either side of the transition.

The $B$-$C_{b}$ transition appears when we keep $\kappa_0$ fixed (and not too large) and, coming from inside $C_{b}$, 
decrease the coupling $\Delta$. 
Its pseudo-critical value $\Delta_c (N_4)$ is a function of the system size $N_4$. By studying its behaviour as a function of 
$N_4$ we have found a dependence which can be fitted well to the functional form
\beql{jnew1}
\Delta_c(N_4) =\Delta_c(\infty ) - c^2  N_4^{-1/\gamma}, 
\eeq
with some non-vanishing constant $c$ and an exponent $\gamma\! \approx\!  2.4$ that within measuring accuracy agrees with the corresponding exponent 
$\gamma \! =\! 2.51(3)$ determined originally 
for a system with large time period \cite{samo}. 

\begin{figure}
\centering
\scalebox{0.25}{\includegraphics{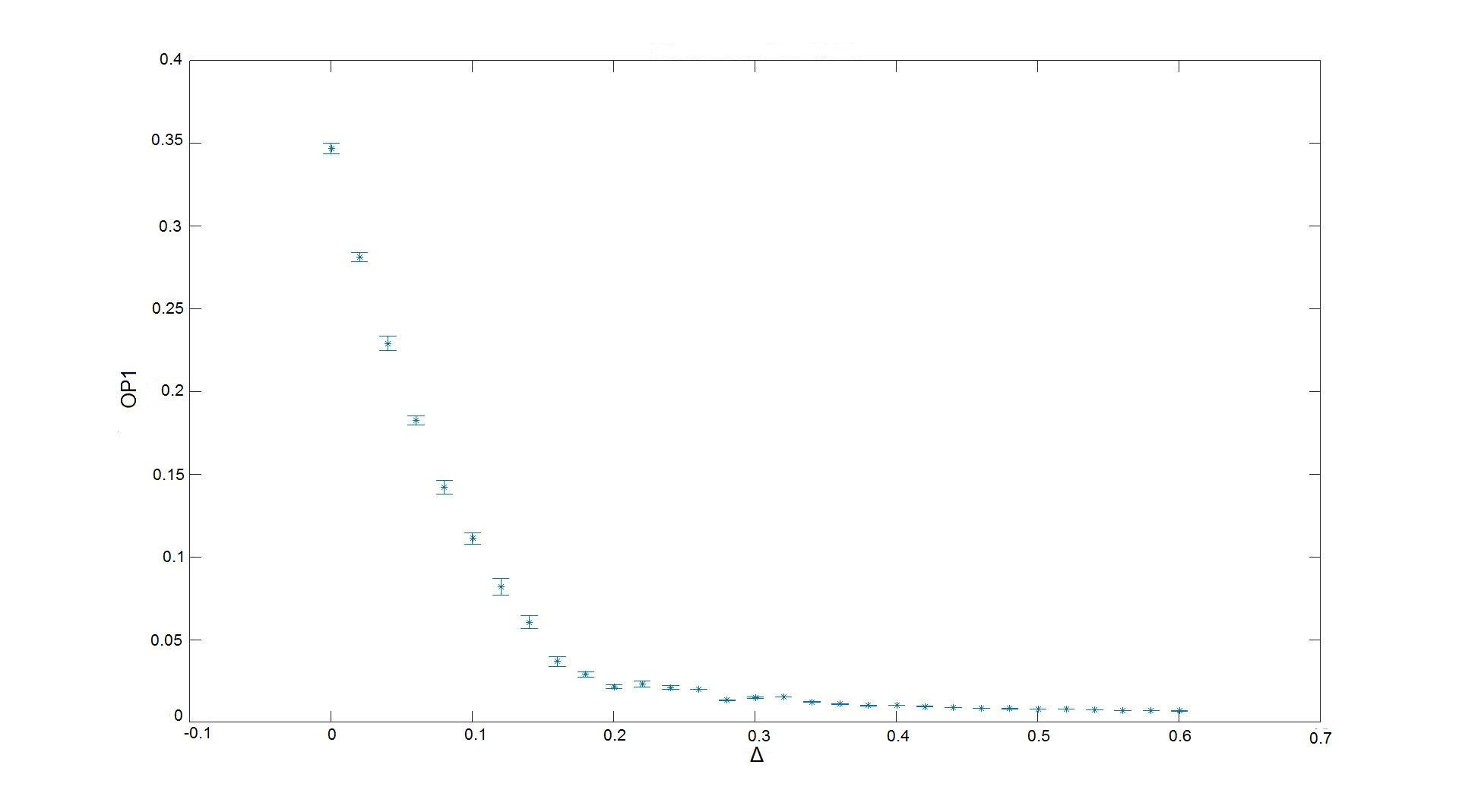}}
\caption{The order parameter $OP_1$ as function of the coupling $\Delta$, measured at $t_{tot}\! =\! 2$, $N_4\! =\! 10k$ and $\kappa_0\! =\! 2.2$,
indicating the presence of a phase transition between the de Sitter and bifurcation phases.}
\label{nilas4}
\end{figure}

In a similar vein, one can compare the behaviour of order parameters away from the 
$B$-$C_{b}$ transition, into phase $C_{b}$ and beyond, by increasing $\Delta$ for fixed $\kappa_0$.
As an example, Fig.\ \ref{nilas4} shows the behaviour of the
order parameter $OP_1$, defined as the absolute value of the difference of
the average spatial curvatures of two adjacent spatial slices,
\begin{equation}
OP_1:=| {\bar R}(t)-{\bar R}(t+1)|, \;\;\; {\bar R}(t)=2 \pi \frac{N_0(t)}{N_3(t)}-const.,
\label{op1def}
\end{equation} 
where $N_0(t)$ and $N_3(t)$ denote the numbers of vertices and spatial tetrahedra contained in the 
spatial triangulation at time $t$. 
This quantity is one of several order parameters first introduced in \cite{bifurcation2} to study the newly 
discovered $C_{dS}$-$C_{b}$ phase transition. The data points shown in Fig.\ \ref{nilas4}, measured at $t_{tot}\! =\! 2$, 
are qualitatively very similar 
to measurements of the same quantity for large $t_{tot}$ \cite{bifurcation2,cgj}.\footnote{Another difference is that in
previous work \cite{bifurcation2,cgj} $N_{41}$ was kept constant. However, unlike what happens at the $B$-$C_{b}$ transition,
inside phases $C_{b}$ and $C_{dS}$ and away from this transition the ratio of $N_{41}$ and $N_{32}$ does not change significantly
when $\Delta$ is varied. We therefore do not expect physical results in this region to depend on the type of volume-fixing.} This
holds for the entire range of $\Delta \in [0, 0.6]$ considered here, with the $C_{dS}$-$C_{b}$ phase transition 
presumably located around
$\Delta\! =\! 0.2$. For the volume $N_4\! =\! 10k$ used presently, the $B$-$C_{b}$ transition lies at
$\Delta\! =\! -0.042(2)$ and therefore well outside the measurement range of Fig.\ \ref{nilas4}. 
Note that the $\Delta$-values in the two-slice system with $t_{tot}\! =\! 2$ are systematically lower than those 
of the system with full time extension $t_{tot}\! =\! 80$ of \cite{samo}, including for the extrapolated critical value
$\Delta_c(\infty )$ of the $B$-$C_{b}$ transition. Comparing with the results of \cite{bifurcation2,cgj}, where the
order of the $C_{dS}$-$C_{b}$ transition is analyzed in more detail, 
the same seems to be true for this transition also. This is not surprising, since the systems are
genuinely different and the location of a critical point is not a universal quantity.
We conclude that our simulations with $t_{tot}\! =\! 2$ reproduce the same characteristics of the 
bifurcation phase $C_{b}$ and the adjacent phase transitions as were already seen for the large-time system with $t_{tot}\! =\! 80$.
The two-slice system therefore seems well suited for a further investigation of this phase.

\subsection{Singular vertices}
\label{singvert}

A key feature of the bifurcation phase, already reported in \cite{bifurcation2}, is the 
appearance of a single ``singular" vertex\footnote{Strictly speaking, there is nothing singular about these vertices
from the point of view of piecewise linear geometry at finite volume. We will nevertheless stick
with this notion, which was originally coined in the context of Euclidean Dynamical Triangulations \cite{singular}.} 
of very high coordination number (this is the number $n_c(v)$ of four-simplices sharing a vertex $v$)
on every second spatial slice. Coming from the de Sitter phase and moving into the bifurcation phase by lowering $\Delta$, 
one finds that a gap opens between the coordination number of the vertex with largest $n_c$ and
that of the vertex with the second-largest $n_c$. Well inside phase $C_{b}$, the maximal $n_c(v)$ in a spatial slice containing such 
a singular vertex is typically orders of magnitude bigger than the average coordination number in the slice. 
At the same time, such a vertex is also singular from a purely three-dimensional point of view, in the sense that it 
is also shared by an exceptionally large number of spatial tetrahedra inside the spatial slice
where it is located.
Another observation, made in \cite{bifurcation2}, is that in simulations with large $t_{tot}$ and therefore many spatial slices, the singular vertices on
alternating slices are associated with a four-dimensional substructure of the triangulation, which takes the form of
a chain of ``diamond-shaped" regions in the time direction. This substructure is imbedded in the rest of the triangulation and
contains a large, finite fraction of the triangulation's total four-volume. 

As already remarked in \cite{bifurcation2}, the presence in the bifurcation phase $C_{b}$ of singular vertices and the structures 
associated with them breaks the homogeneity and isotropy (on average) of geometry which is present in the de Sitter phase $C_{dS}$. 
Given the way Causal Dynamical Triangulations are implemented, there is nothing in principle that prevents homogeneity and isotropy 
of the average universe modelled by CDT triangulations, in the limit as the lattice spacing is taken to zero. 
This is indeed what is observed in phase $C_{dS}$, where a number of properties of the dynamically generated
``quantum universe" are very well described by a minisuperspace model with built-in spatial homogeneity and isotropy \cite{semi, CDT1}.
More than that, in $C_{dS}$ the average shape of the universe can be fitted to a de Sitter space, a maximally symmetric spacetime
solving the classical Einstein equations.  
The appearance of isolated vertices of very high coordination number in phase $C_{b}$ is clearly incompatible with these symmetries.
Given that phase transitions in physical systems are often related to the breaking of a symmetry, it is natural to associate the 
$C_{dS}$-$C_{b}$ phase transition with a symmetry breaking also, namely, of homogeneity and isotropy. 

\subsection{Singular vertices cause bifurcation split}

In what follows, we will provide further 
evidence that phase $C_{b}$ is associated with the appearance of 
singular vertices and that they can be viewed as the decisive characteristic of the bifurcation phase. 
More specifically, we will establish a quantitative relation between the ``bifurcation split", the observed 
typical volume difference between neighbouring spatial slices \cite{bifurcation2,cgj}, and the order of the singular vertex present.
We will set $\Delta\! =\! 0$, which for the volumes considered places us in the bifurcation phase, and at a safe
distance from either of the adjoining phase transitions. 

\begin{figure}[t]
\centering
\scalebox{0.75}{\includegraphics{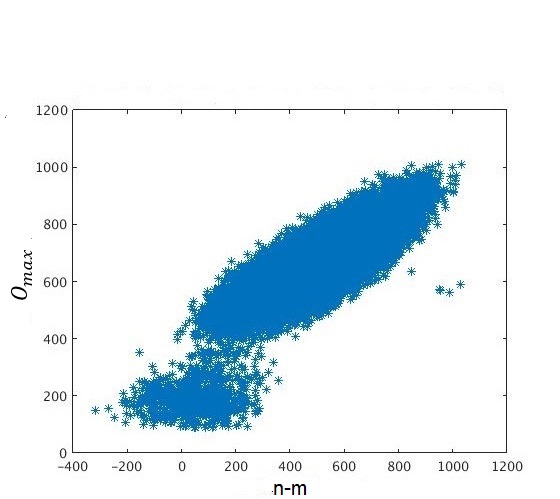}}
\caption{Distribution of the highest vertex order $O_{max}$ versus the difference $n-m$ of the spatial slice volumes, where by definition
the highest-order vertex is contained in the slice of volume $m$. Data taken in the bifurcation phase ($\kappa_0\! =\! 2.2$, $\Delta\! =\! 0$,
$N_4\! =\! 10k$).
}
\label{scatter}
\end{figure}
To analyze the geometry of the triangulations with $t_{tot}\! =\! 2$ in greater detail, 
we will use a variant of the notion of vertex order, which for a given vertex $v$
counts the number of (4,1)-simplices between the two slices that share the vertex $v$ and have a spatial three-simplex 
in common with the spatial slice {\it not} containing $v$. Using this definition\footnote{We have checked that other notions
of vertex order, including the coordination number $n_c$ defined in subsection (\ref{singvert}), lead to equivalent results. 
The vertex order used presently is convenient since it is directly related to the diamond volume.},
we will call $O_{max}$ the maximal vertex order occurring in
a given two-slice configuration. When a singular vertex is present, $O_{max}$ will coincide with the order of this vertex.
Like in our earlier discussion of the matrix elements (\ref{trama}) of the reduced transfer matrix, we will use the letters $m$ and $n$
to denote the three-volumes of the two adjacent spatial slices. 
In addition, by definition, $m$ will denote the volume of the slice that contains the vertex of maximal order, 
and $n$ the volume of the slice that does not. 
Note that if a singular vertex $v_s$ is present in the spatial slice of volume $m$, 
$O_{max}\leq n$ is the three-volume of the intersection of the (half-)diamond with tip $v_s$ and the spatial slice of volume $n$. 

In Fig.\ \ref{scatter} we show the distribution of the highest vertex order $O_{max}$ versus the volume difference $n\! -\! m$ of the two 
spatial slices. One can roughly distinguish two regions. Below $O_{max}\!\approx\! 300$, the configurations contain no singular vertex 
in the sense that there is no significant gap between $O_{max}$ and the orders of the other vertices. 
A closer analysis reveals that for fixed $O_{max}$ in this region, 
the distribution of the volume differences is approximately Gaussian around $n\! -\! m\! =\! 0$. In other words, 
neighbouring slices preferentially have equal volumes. From previous investigations \cite{bifurcation1} we recognize this latter property 
as characteristic for configurations inside
the de Sitter phase $C_{dS}$. These configurations by no means dominate the dynamics of the bifurcation phase 
studied here, but the system makes occasional 
excursions to them, at least for the spacetime volume we are considering. This will be further corroborated by data presented
below. 
\begin{figure}
\centering
\scalebox{0.32}{\includegraphics{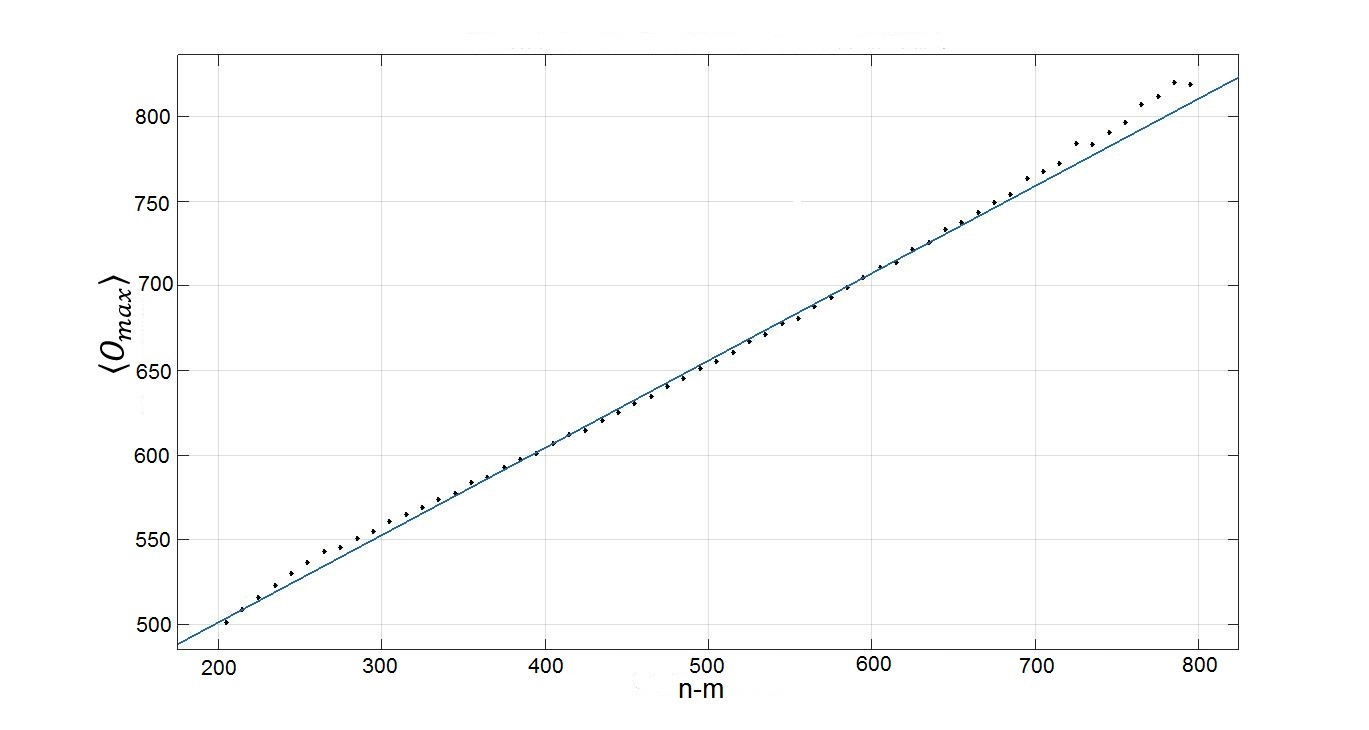}}
\caption{Expectation value of the highest vertex order $O_{max}$ 
as function of the difference $n\! -\! m$ of the spatial slice volumes (same specifications as in Fig.\ \ref{scatter}).}
\label{nilas14}
\end{figure}

The vast majority of configurations lie in the region where $O_{max}\!\gtrsim\! 400$. Around $O_{max}\! =\! 400$ a gap opens
between $O_{max}$ and the distribution of the orders of the remaining vertices that becomes larger as the value of $O_{max}$ increases,
signalling the appearance of a singular vertex. At the same time, at fixed $O_{max}$,
the configurations are now peaked around a nonvanishing volume difference.\footnote{Of course, these statements 
should be understood as statistical statements, arrived at by analyzing many configurations.} 
This is typical for the bifurcation phase $C_{b}$, where
the effective transfer matrix $\langle n |M|m\rangle$ has a double-peak structure as function of the volume difference $n\! -\! m$ (and at
fixed $m\! +\! n$), unlike the single peak found in $C_{dS}$. It entails that the two slice volumes preferentially differ by 
a finite amount $\langle |n-m|\rangle \not= 0$. 

The interesting new finding from our data is that the    
expectation value $\langle O_{max}\rangle$ depends linearly on
this ``bifurcation split" $n-m$ between the two spatial volumes, where again the slice with the lower volume $m$ is the one
containing the highest-order vertex. This linear relation is illustrated in Fig.\ \ref{nilas14}.
Extrapolating $ n\! -\! m $ down to zero one obtains a vertex order of around 400, 
in agreement with Fig.\ \ref{scatter}. 
We conclude that the bifurcation phenomenon, observed in previous studies of the effective transfer
matrix \cite{bifurcation2}, seems to be a function of the appearance of singular vertices. 

The particular choice of coupling constants for which the above results have been obtained 
is associated with specific expectation values for both the highest vertex order $O_{max}$ and
the bifurcation split $n\! -\! m$. Not surprisingly, these variables have Gaussian-like distributions
around their mean values. For example, $O_{max}$ has an approximate Gaussian distribution 
peaked at 675 with standard deviation around 50. Although it is not very visible on the scatter
plot of Fig.\ \ref{scatter}, there are therefore many fewer configurations with vertex order 500 or 900, say, than
there are with vertex order 700. Furthermore, for each given value of $O_{max}$ the width of the (Gaussian) 
distribution of $n\! -\! m$ is approximately the same and coincides with the one determined by the effective 
action associated with the effective transfer matrix. 
This implies that the width is not a function of the vertex order for fixed values of the coupling constants.
 
\begin{figure}
\centering
\scalebox{0.8}{\includegraphics{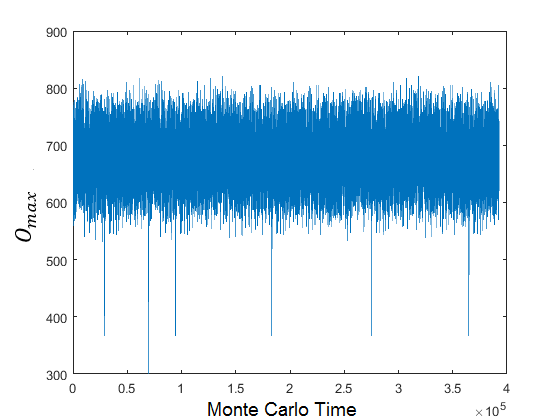}}
\caption{Time series of the maximal vertex order $O_{max}$ as a function of Monte Carlo time, exhibiting rare dips to values below 500.}
\label{nilastimeseries}
\end{figure}

The much rarer configurations with $O_{max}\lesssim 400$ have a special status, a fact that becomes clear
when studying the maximal vertex order as a function of Monte Carlo time. As shown in Fig.\ \ref{nilastimeseries},
$O_{max}$ fluctuates around 675. Since there is a gap in the vertex order distribution below the maximal value, 
and since vertex orders can only change by relatively small amounts in each Monte Carlo update, 
the highest-order vertex usually remains located firmly in one of the two spatial slices. 
However, occasionally $O_{max}$ takes a very fast dip to a value below 500, which means that the
distinguished, singular vertex disappears. After such a dip, a new singular, highest-order vertex appears
randomly on either one of the
spatial slices. We do not yet understand in detail how this process works, but the excursions occur seldom and their durations are much 
too short in Monte Carlo time
to be explained as random processes associated with the Gaussian distribution of $O_{max}$.
The configurations with $O_{max}\!\leq\! 500$ in Fig.\ \ref{scatter} constitute less than 0.1\% of the total number of configurations.
 
\begin{figure}
\centering
\scalebox{0.30}{\includegraphics{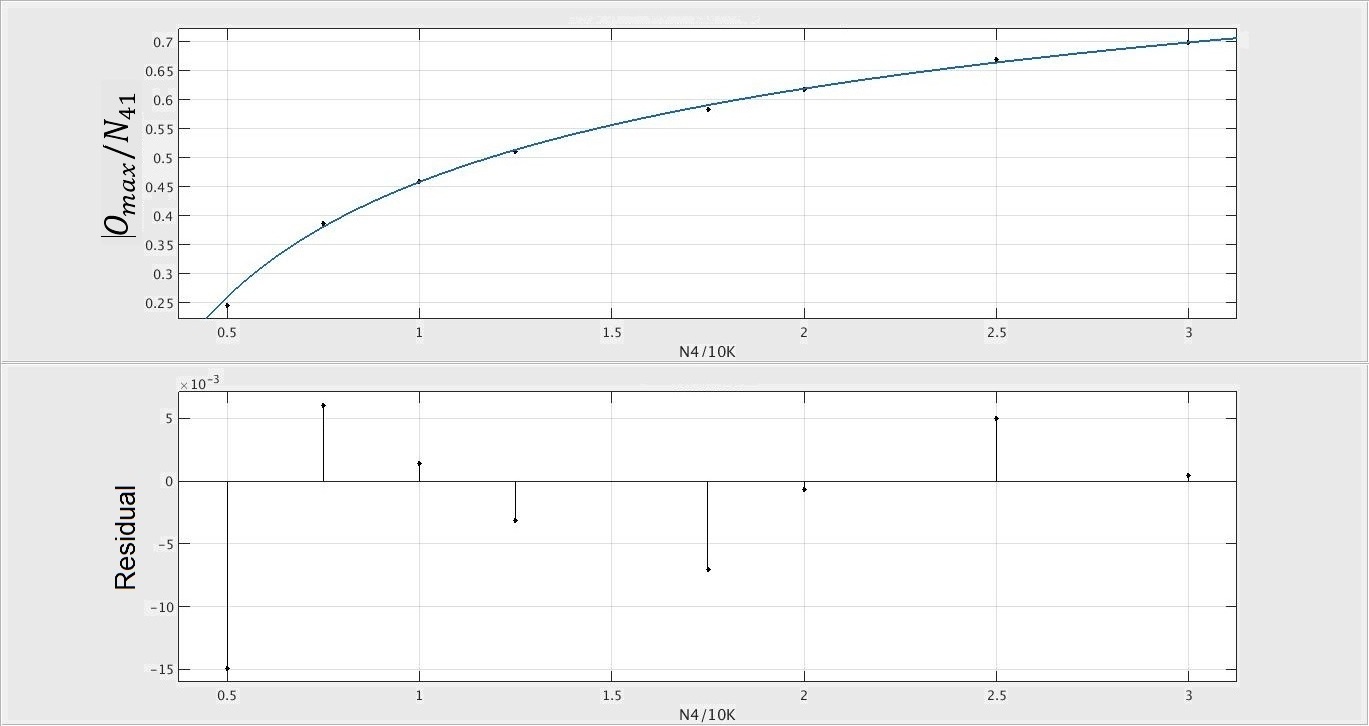}}
\caption{Relative order $O_{max}/N_{41}$ of the singular vertex as a function of $N_4$, at 
$\Delta\!  =\! 0$, together with a fit $-0.9 x^{-0.3} + 1.3$, $x = N_4/10.000$, and corresponding residuals. 
Note that the fit cannot be entirely accurate for large volumes, because for 
$x\gtrsim 39$ it gives values larger than 1, which is not permissible.}
\label{nilas8}
\end{figure}

Finally, we would like to understand whether there is just one singular vertex in a given spatial slice or whether further
vertices with exceptionally high order can appear in the same slice when the system size goes to infinity. 
Studying $O_{max}$ as a function of the total spacetime volume $N_4$, we found that
the relative order of the singular vertex, that is, $O_{max}$ divided by $N_{41}$, 
grows with $N_4$, as shown in Fig.\ \ref{nilas8}. 
Recall that $O_{max}$ coincides with the four-dimensional volume of the (half-)diamond whose tip is the singular vertex, 
which in turn is bounded by $N_{41}$. Since the measured ratio $O_{max}/N_{41}$ is a finite fraction of 1, there can be at most 
a finite number of similarly ``singular" vertices in the limit $N_4 \to \infty$. Presumably it is just the single singular vertex on every second spatial slice 
we see at lower volumes. 
However, the detailed interpretation of this infinite-volume limit requires some care. 
The point is that just taking $N_4 \to\infty$ at fixed coupling 
constant $\Delta$ corresponds to changing the real, effective coupling constant. 
The presence of such a volume dependence is apparent from eq.\ \rf{jnew1}, which describes how the pseudo-critical 
$\Delta_c(N_4)$ increases with increasing $N_4$.
In the case at hand, for sufficiently large volume $N_4$ (larger than what we have considered here), our
present choice $\Delta\! =\! 0$ will therefore no longer lie in phase $C_{b}$, but in phase $B$. 
With this caveat in mind, our data indicate that in the infinite-volume limit, the CDT ensemble with $t_{tot}\! =\! 2$
contains just one singular vertex.

\section{Summary and conclusion}
\label{final}

In this paper we have investigated the bifurcation phase $C_{b}$ recently discovered in CDT quantum gravity. 
We first re-examined the $B$-$C_{b}$ phase transition (formerly called the $B$-$C$ transition). The order parameter
used previously to determine the order of this transition exhibited an unexpected dependence on how 
the total spacetime volume was fixed in the simulations: keeping the total number $N_4$ of four-simplices fixed resulted
in a double-peak distribution for the order parameter, whereas keeping the number $N_{41}$ of four-simplices
of type (4,1) fixed yielded only a single peak. 
A careful examination of the entropy factor $\cN (N_{41}, N_{32}, N_0)$ revealed
that in the volume range considered it has a rather complicated form as a function of $N_{41}$ and $N_{32}$,
which completely explains the observed behaviours of the order parameter for the two different volume fixings.
These findings reconfirm that the double-peak structure seen for $N_4\! =\!  const$ in no way contradicts the
earlier conclusion that the $B$-$C_{b}$ phase transition is of second order \cite{samo}.

The fact that the new $C_{dS}$-$C_{b}$ transition 
was discovered in simulations with a short total time extension $t_{tot}$, to determine the so-called effective transfer matrix,
raised the question of whether the choice of $t_{tot}$ (as a long or short compactified time direction) has an influence on the observed phase structure. 
In the measurements presented above we have not found any indication that this is the case. The $B$-$C_{b}$ transition is
still present for the system with $t_{tot}\! =\! 2$, with a signal compatible with that observed for $t_{tot}\! =\! 80$. 
Earlier work \cite{bifurcation2,cgj} had already shown that the new $C_{dS}$-$C_{b}$ transition between the de Sitter and
the bifurcation phase is also present for large $t_{tot}$, and clearly visible for appropriate choices of order parameters. 
There are preliminary indications that this transition could be of higher order too \cite{cgj}, but more extensive simulations are needed 
to obtain more conclusive results. 

The equivalence between long and short $t_{tot}$ motivated our further study of the properties of the bifurcation phase 
by considering the somewhat simpler two-slice system.
We showed that the behaviour of the highest-order, ``singular" vertex that appears in this phase is directly related to 
the previously observed tendency of the neighbouring spatial slices to develop a nonvanishing mean volume difference
or ``bifurcation split". More specifically, the maximal vertex order $O_{max}$ scales linearly with this volume difference.
This gives us a more detailed, geometrical understanding of the mechanism behind the bifurcation split: a finite fraction
of the (4,1)-simplices between the two spatial slices clusters into a half-diamond whose tip is the singular vertex. This half-diamond
forms a substructure, which is imbedded in the rest of the triangulation and leads to a corresponding ``excess" of three-volume of the slice
{\it not} containing the singular vertex. 

At the same time, the appearance of a singular vertex\footnote{or of $t_{tot}/2$ singular vertices when working with a larger 
(even) number $t_{tot}>2$ of spatial slices} when crossing into phase $C_{b}$ from
the de Sitter phase signals
a breaking of the homogeneity and isotropy of geometry present in the de Sitter phase on scales above the cutoff scale.
It suggests that the bifurcation-de Sitter phase transition can be associated with the breaking of a symmetry, a situation
common in non-geometric statistical systems. 

From this point of view the $C_{dS}$-$C_{b}$ phase transition resembles the phase transition between 
the branched-polymer and the crumpled phase in (Euclidean) Dynamical Triangulations.
The DT configurations in the branched-polymer phase appear homogenous and isotropic
(although not in any sense that is associated with a four-dimensional spacetime), 
while configurations in the crumpled phase are characterized by the appearance of two distinguished,
singular vertices of very high order and a 
singular link in between them \cite{singular}. 
Unfortunately, in this purely Euclidean quantum gravity model 
the phase transition between the two phases is only a first-order transition, even in extended DT models
with an additional coupling constant, as already mentioned in the Introduction. 
In CDT we may be in the more exciting situation that
the analogous $C_{dS}$-$C_{b}$ phase transition is of second order, like the $B$-$C_{b}$ transition,
and therefore may be used to define a continuum theory of quantum gravity.

From a more general perspective, our investigation has given us additional insights into the type of
mechanisms that can drive the nonperturbative dynamics of systems of (a priori) higher-dimensional 
geometry and the appearance of phase transitions, our understanding of which is rather limited.
A conclusion we can already draw at this stage is that the phase structure of Causal Dynamical
Triangulations in four dimensions, despite
the presence of only two tuneable bare parameters, is amazingly rich 
and presents us with further opportunities to uncover viable continuum theories of
quantum gravity.

\vspace{.5cm}

\noindent {\bf Acknowledgements.} 
JGS and JJ wish to acknowledge the support of grant DEC-2012/06/A/ST2/00389
from the National Science Centre Poland.

\section*{Appendix}
In this appendix we construct a simple model function ${\cal F}(N_{41}, N_{32})$ for the free energy 
$F(N_{41}, N_{32})$ introduced in eq.\ (\ref{eq:F}), which reproduces the single- and double-peak
signals at the $B$-$C_{b}$ phase transition described in Sec.\ \ref{puzzle}.
It is based on an ansatz of the form
\begin{equation}
{\cal F}(N_{41}, N_{32}) = c_1 +g_1(N_{41}) + g_2(N_{32}) + c_2\, g_3(N_{41}) g_4(N_{32}),
\label{eq:Fg}
\end{equation}
where the $c_i$ are constants and the $g_i$ are functions of the counting variables $N_{41}$ and $N_{32}$
as indicated. The motivation behind this ansatz is that it can in principle account in a simple way
for the observed single- and double-peak structure of the probability distributions of the parameter
$x\! =\! N_{41}\! -\! N_{32}$, as follows. Assume that all $g_i$ are quadratic functions of their
arguments in a reasonably large part of the $(N_{41},N_{32})$-plane 
displayed in Fig.\ \ref{fig:F}. For fixed $N_{41}$, ${\cal F}$ is a quadratic function of $N_{32}$
and therefore of $x$, explaining the shape of the 
observed probability distribution $\tilde P(x)$ shown in Fig.\ \ref{fig:distN41}.
Conversely, for fixed $N_{4}\! =\! N_{41}\! +\! N_{32}$, the model function $\cal F$ will be a
fourth-order polynomial in $x$, and can in principle account for the double peak in the 
observed probability distribution $\bar P(x)$ depicted in Fig.\ \ref{fig:distN4}. 

On the other hand, it is clear that the ansatz (\ref{eq:Fg}) with quadratic functions $g_i$ cannot
be the whole story, because it would make the associated probability distribution $\tilde P(x)$,
obtained by exponentiating $\cal F$ according to eq.\ (\ref{jnew2}), a pure Gaussian. 
However, the standard deviation calculated using \rf{ju1} would then be independent 
of the coupling $\kappa_{32}$, and could not give rise to a peak like the ones shown in Fig.\ \ref{fig:deviation}. 
The fact that there is an ansatz which produces a double peak in $\bar P(x)$, 
but no signal of critical behaviour in $\tilde P(x)$ further corroborates the original statement in \cite{samo} 
that the presence of a double peak is not per se related to a (first-order) transition.

We will now show that using the ansatz (\ref{eq:Fg}) in the rectangular region $N_{41}\in [30.000,37.000]$, $N_{32}\in [3.000,11.000]$ 
and imposing suitable normalization conditions on the functions $g_i$, they can be
determined uniquely from the data, without assuming any specific functional form for them. 
Also the constants $c_i$ get determined uniquely.
Substituting the extracted functions and constants into (\ref{eq:Fg}) yields a function ${\cal F}(N_{41}, N_{32})$
that agrees with the directly measured ${F}(N_{41}, N_{32})$ up to noise in the data. 

The additional constraints we impose are
\begin{eqnarray}
\label{constr}
	&\langle g_1(N_{41}) \rangle_{N_{41}} \! =\! \langle g_2(N_{32}) \rangle_{N_{32}}\! =\! 
	 \langle g_3(N_{41}) \rangle_{N_{41}}\! =\! \langle g_4(N_{32}) \rangle_{N_{32}} = 0,\nonumber\\
	 &\langle g_3(N_{41})^2 \rangle_{N_{41}} \! =\! \langle g_4(N_{32})^2 \rangle_{N_{32}} = 1,
\end{eqnarray}	 
where for each of the two variables $z=N_{41},N_{32}$ the average $\langle f(z) \rangle_z$ of a function $f(z)$ 
is defined as
\begin{equation}
\langle f(z) \rangle_z  := \frac{1}{z_{max}\! -\! z_{min}\! +\! 1}\; \sum_{z=z_{min}}^{z_{max}} f(z). 
\end{equation}
\begin{figure}
\centering
\scalebox{1.0}{\includegraphics{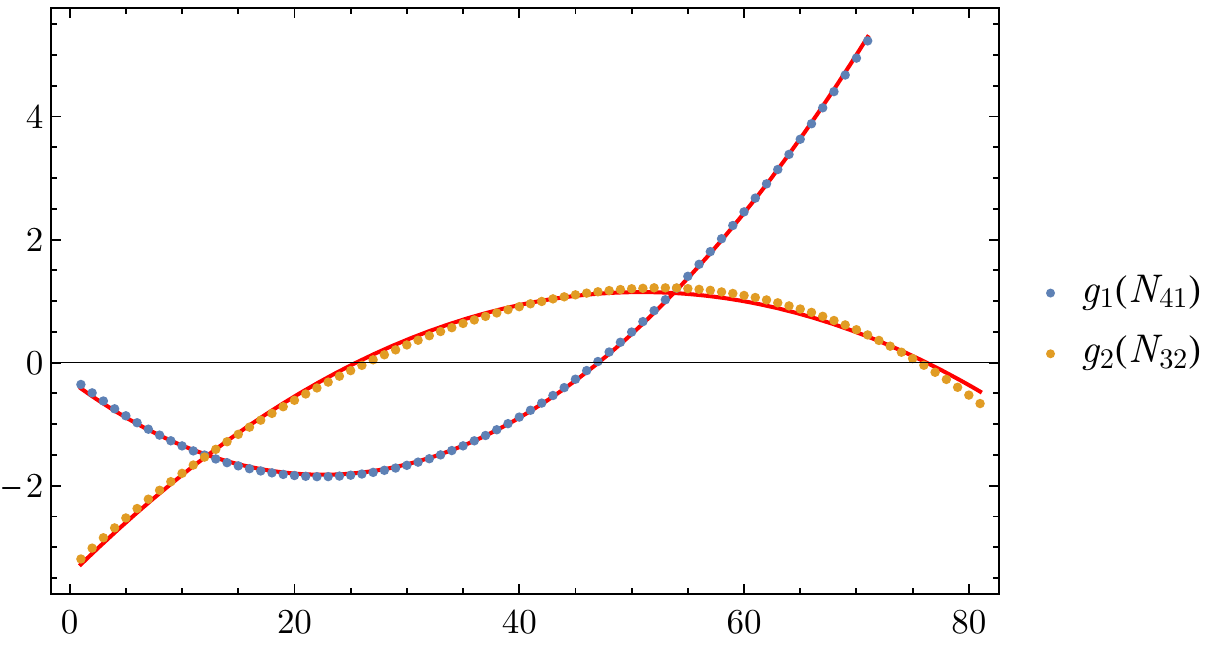}}
	\caption{The functions $g_1(N_{41})$ (blue dots) and $g_2(N_{32})$ (yellow dots) of the ansatz (\ref{eq:Fg}), extracted from 
	the measured free energy $F(N_{41},N_{32})$. The ranges of the parameters $N_{41}$ and $N_{32}$ on the horizontal axis have
	been rescaled by a common factor and shifted to fit them into a single coordinate system.
	The red curves are best fit quadratic functions.}
	\label{fig:g1g2}
\end{figure}
Taking into account the relations (\ref{constr}), three of the unknown quantities can be constructed directly from the
measured function $F(N_{41},N_{32})$, namely,
\begin{eqnarray}
	c_1 &=& \langle F(N_{41}, N_{32})\rangle_{N_{41},N_{32}}, \nonumber\\
	g_1(N_{41}) &=& \langle F(N_{41}, N_{32})\rangle_{N_{32}} - c_1,\\
	g_2(N_{32}) &=& \langle F(N_{41}, N_{32})\rangle_{N_{41}} - c_1 ,\nonumber
\end{eqnarray}
where the first equation involves a double average.
We can find the remaining functions $g_3(N_{41})$ and $g_4(N_{32})$ by solving an eigenproblem.
Let us define the two matrices
\begin{eqnarray}
	\Phi_{N_{41}, N_{32}} &:=& F(N_{41}, N_{32}) - c_1 - g_1(N_{41}) - g_2(N_{32}),\\
	{\cal M}_{N_{41}, N_{32}} &:= & \Phi_{N_{41}, N_{32}} - c_2 \ g_3(N_{41})\  g_4(N_{32}). 
\end{eqnarray}
To find the best approximation of $F(N_{41}, N_{32})$ by ${\cal F}(N_{41}, N_{32})$ we have to minimize 
the error function $E$, defined as
\begin{equation}
\label{mm}
E = \sum_{N_{41}, N_{32}} \left( {\cal F}(N_{41}, N_{32}) - F(N_{41}, N_{32})\right)^2=
\sum_{N_{41}, N_{32}} {\cal M}^2_{N_{41},N_{32}} = \Tr \, {\cal M}{\cal M}^T.
\end{equation}
One can show that in order to extremize (\ref{mm}), $g_3(N_{41})$ must be an eigenvector of the
matrix $\Phi \Phi^{T}$, and $g_4(N_{32})$ an eigenvector of the matrix $\Phi^{T} \Phi$.
To minimize $E$, one must choose the eigenvectors corresponding to the largest eigenvalues,
a condition that fixes $g_3(N_{41})$ and $g_4(N_{32})$. The largest eigenvalue is positive and
has the same value $c_2^2$ for both matrices, which also fixes the (positive) constant $c_2$.

Having determined all functions $g_i$ and constants $c_i$,
the resulting model function ${\cal F}(N_{41}, N_{32})$ differs from the original empirical function $F(N_{41},N_{32})$ 
only by what looks like noise, which suggests that in the selected region the functional form assumed in (\ref{eq:Fg}) is very accurate.

The extracted functions $g_1(N_{41})$ and $g_2(N_{32})$ are shown in Fig.\ \ref{fig:g1g2}.
Although $g_1(N_{41})$ is approximated very well by a quadratic function in a neighbourhood around its
minimum, that is, in the range of $N_{41}$ we have been considering, this cannot possibly be true for
its entire range. The reason is that a quadratic dependence would imply an entropy growth
proportional to $e^{+\const \cdot N^2_{41}}$, which would contradict the fact, proven in \cite{dj},
that the number of triangulations can grow at most exponentially with $N_{41}$. 
The quadratic fit displayed in Fig.\ \ref{fig:g1g2} must therefore be a local approximation arising as
an expansion of a slower growing function of $N_{41}$.

\begin{figure}
\centering
\scalebox{1.0}{\includegraphics{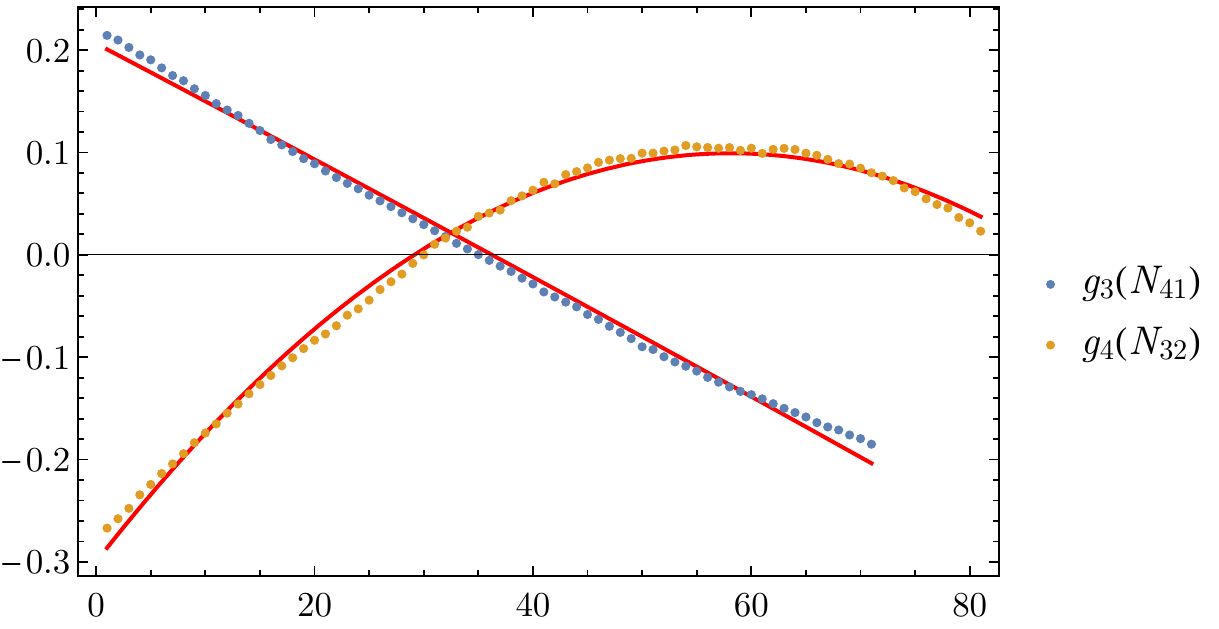}}
\caption{The functions $g_3(N_{41})$ (blue dots) and $g_4(N_{32})$ (yellow dots) of the ansatz (\ref{eq:Fg}), extracted from 
	the measured free energy $F(N_{41},N_{32})$ as described in the text. 
Again, the ranges of the parameters $N_{41}$ and $N_{32}$ on the horizontal axis have
	been rescaled and shifted.
	The red curves are best fit quadratic functions. 
	}
\label{fig:g3g4}
\end{figure}

The extracted functions $g_3(N_{41})$ and $g_4(N_{32})$ are shown in Fig.\ \ref{fig:g3g4}.
Because both $g_2(N_{32})$ and $g_4(N_{32})$ are well approximated by quadratic polynomials, 
the corresponding distribution $\tilde{P}(x)$ for fixed $N_{41}$ (c.f. eq.\ (\ref{jnew2})) is almost Gaussian.
The function $g_3(N_{41})$ is nearly linear so $g_3(N_{41}) \cdot g_4(N_{32})$
results in a decreasing width of the distribution ${\cal P}(N_{41}, N_{32})$ as $N_{41}$ grows
(the larger $N_{41}$, the more negative is the coefficient in front of $N^2_{32}$ in $F(N_{41}, N_{32})$). 
Going back to our earlier Fig.\ \ref{fig:FN4} depicting the distribution $\bar P(x)$ of $x$ for fixed $N_4$,
the green curve is based on the ansatz \rf{eq:Fg} with quadratic functions 
$g_i$.\footnote{To improve the quality of the fit, the quadratic functions we used are slightly different from the quadratic functions
shown in Figs.\ \ref{fig:g1g2} and \ref{fig:g3g4}, since most of the region of the 
diagonal line $N_4\! =\! 40k$ in Fig.\ \ref{fig:F} used in the fit is outside the rectangular region 
used to determine the functions shown in Figs.\ \ref{fig:g1g2} and \ref{fig:g3g4}.} It fits the data based 
directly on the measured free energy $F(N_{41},N_{32})$ (indicated
by the blue dots) quite well.

To summarize, we have demonstrated that a simple ansatz like \rf{eq:Fg}, with quadratic functions $g_i(x)$
can reproduce the observed features of the probability distributions $\tilde P(x)$ and $\bar P(x)$,
without at the same time reproducing any signal of critical behaviour. This further supports earlier
assertions that the appearance of a double peak in $\bar P(x)$ is not necessarily related to
any specific form of critical behaviour.

\end{document}